\documentstyle[11pt]{article}

\newcommand{\half}{\frac{1}{2}}
\newcommand{\cir}[1]{n_{#1}}
\newcommand{\Ztwo}{({\bf Z}_2)^{\nu-2}}

\newcommand{\A}[1]{{\cal A}_{#1}}
\newcommand{\B}[1]{{\cal B}_{#1}}
\newcommand{\bra}[2]{\left\{ \begin{array}{c} #1 \\ #2 \end{array} \right\}}
\newcommand{\foot}[2]{\mbox{\begin{tabular}{c} {\scriptsize #1}
\\ {\scriptsize #2} \end{tabular}}}
\newcommand{\r}{\hat{r}}
\newcommand{\s}{\hat{s}}
\newcommand{\n}{\tilde{n}}
\newcommand{\sumt}{\widetilde{\sum}}
\newcommand{\sumtt}{\widetilde{\widetilde{\sum}}}

\newtheorem{lemma}{Lemma}

\begin{document}

\title{Polynomial Fermionic Forms for the Branching Functions of the Rational
Coset Conformal Field
Theories $\widehat{su}(2)_{M}\times \widehat{su}(2)_{N}/\widehat{su}(2)_{M+N}$}
\author{Anne Schilling \thanks{e-mail address: anne@max.physics.sunysb.edu}\\
\footnotesize{{\em Institute for Theoretical Physics,}}\\
\footnotesize{{\em State University of New York at Stony Brook, Stony Brook,
NY 11794-3800}}}

\date{}

\maketitle

\begin{abstract}
General fermionic expressions for the branching functions of the
rational coset conformal field theories $\widehat{su}(2)_{M}\times
\widehat{su}(2)_N/\widehat{su}(2)_{M+N}$ are given. The equality of the
bosonic and fermionic representations for the branching functions is proven by
introducing polynomial truncations of these branching functions which are
the configuration sums of the RSOS models in regime III. The path space
interpretation of the RSOS models provides recursion relations for the
configuration sums. The proof of the recursion relations for the fermionic
expressions is given by using telescopic expansion techniques.
The configuration sums of the RSOS model in regime II which correspond to
the branching functions of the $Z_{M+N}$-parafermion conformal field theory
are obtained by the duality transformation $q\rightarrow q^{-1}$.
\end{abstract}

\section{Introduction}
\label{sec-intro}

Recently there has been much interest in the explicit representations
as fermionic
$q$-series of the characters and branching functions of conformal
field theories and many general fermionic forms have been found/conjectured
for the coset models $(G_l)_m\times (G_l)_n/(G_l)_{m+n}$ where $G$ is a
simply laced Lie algebra (and the levels are integer) and the generalization
to fractional levels of the minimal models $M(p,p')$
and the $N=1$ superconformal models $SM(p,p')$ \cite{Lepowsky}-\cite{Foda}.
The fermionic
representations are complementary to the already known bosonic
representations \cite{RC}-\cite{Date1}.
The two representations are related
by generalized Rogers-Ramanujan identities.

In particular, Kedem et al. \cite{Kedem} have conjectured the
characters for the coset models
$\widehat{su}(2)_{M}\times \widehat{su}(2)_1/\widehat{su}(2)_{M+1}$
and the identity character for the general coset model
$\widehat{su}(2)_{M}\times \widehat{su}(2)_{N}/\widehat{su}(2)_{M+N}$.
The Virasoro characters for the unitary superconformal minimal models
which correspond
to the coset constructions
$\widehat{su}(2)_{M}\times \widehat{su}(2)_{2}/\widehat{su}(2)_{M+2}$
have been conjectured by Baver and Gepner \cite{Gepner}. Nakayashiki
and Yamada \cite{NY} discuss
the coset models
$\widehat{su}(2)_{M}\times \widehat{su}(2)_{N}/\widehat{su}(2)_{M+N}$
from the crystalline spinon
basis point of view.

The proofs of the fermionic representations which have been given
almost all follow either the method of Schur \cite{Schur} or combinatorical
methods. Schur's method is based on finding a class of polynomials
which satisfy recursion relations
in terms of the degree and in the limit of infinite degree equal the
characters/branching functions. It should be emphasized that the polynomial
truncations of the characters/branching functions are not unique.
Melzer \cite{Melzer} has conjectured a polynomial form for the coset models
$\widehat{su}(2)_{M}\times \widehat{su}(2)_1/\widehat{su}(2)_{M+1}$
which has been proven by Berkovich \cite{Berkovich} via telescopic expansion
techniques and Warnaar \cite{Warnaar}
by counting weighted paths based on the Fermi gas picture. Further such
polynomial recursion
relation proofs have recently been given for
the $N=1$ superconformal models $SM(2,4\nu)$ by Berkovich, McCoy and
Orrick \cite{Will}. Bose-Fermi identities for the coset models
$\widehat{sl}(n)_1\times \widehat{sl}(n)_1/\widehat{sl}(n)_2$ have been proven
by Foda, Okado and Warnaar \cite{Foda} using purely combinatorical methods.

The purpose of this paper is to extend these studies by 1)
presenting a complete set of fermionic branching function formulas for
the coset
models $\widehat{su}(2)_{M}\times \widehat{su}(2)_N/\widehat{su}(2)_{M+N}$
and 2) to prove them via Schur's method by introducing polynomial
generalizations which satisfy
the same RSOS recursion relations which were used by Andrews, Baxter and
Forrester \cite{ABF} and the Kyoto group \cite{Date}, \cite{Date1} to
compute the
bosonic form of the branching functions.

Our principal result for point 1) is that the fermionic form of the
branching functions
for the coset models
$\widehat{su}(2)_{\nu-N-1}\times \widehat{su}(2)_{N}/\widehat{su}(2)_{\nu-1}$
is given by
\begin{equation}
\label{bf}
q^{\eta}c_{\r\,\s}^{(l)}=\sum_{m_i\geq 0,{\rm e.o.}\,{\rm restriction}\,
\hat{Q}_{\r,\s,l}}
q^{\frac{1}{4}mCm-\half \hat{A}_{\r,\s,l}m}\frac{1}{(q)_{m_N}}
 \prod_{i=1,i\neq N}^{\nu-2} \left[ \begin{array}{c}
\half (Im+\hat{u}_{\r,\s,l})_i \\ m_i \end{array} \right].
\end{equation}
Here the sum runs over all $m_i\geq 0$ restricted by $\hat{Q}
\in \Ztwo$ (i.e. if $(\hat{Q})_i=0$
then $m_i$ even and if $(\hat{Q})_i=1$
then $m_i$ odd), $I$ is the $(\nu-2)$ dimensional incidence
matrix $(I)_{a,b}=\delta_{a,b+1}+\delta_{a,b-1}$,
$C=2-I$ the $(\nu-2)$ dimensional Cartan matrix of the Lie
algebra $A_{\nu-2}$ and $\hat{A}$ and
$\hat{u}$ are $(\nu-2)$ dimensional vectors with integer
values. The Gaussian polynomials $\left[ \begin{array}{c} n\\ m
\end{array}\right]$ are defined by
\begin{equation}
\left[ \begin{array}{c} n \\ m \end{array} \right]=
\left\{ \begin{array}{ll} \frac{(q)_n}{(q)_m(q)_{n-m}}
& \mbox{if } 0\leq m\leq n \\ 0 & \mbox{otherwise}
\end{array} \right.
\end{equation}
and
\begin{equation}
(q)_n=\prod_{j=1}^{n}(1-q^j).
\end{equation}
There are four sets of solutions for $\hat{A}, \hat{u}$ and $\hat{Q}$ which
all lead to the branching function $c_{\r,\s}^{(l)}$.
$\hat{A}, \hat{u}$ and $\hat{Q}$ are either given by
\begin{eqnarray}
\label{bfdown}
\hat{A}_{\r,\s,l}&=&e_{\s-1}\nonumber\\
\hat{u}_{\r,\s,l}&=&e_{\s-1}+e_{\nu-\r}+e_{l-1}\nonumber\\
\hat{Q}_{\r,\s,l}&=&(\r-1)\rho+(e_{\s-2}+e_{\s-4}+\ldots)
+(e_{l-2}+e_{l-4}+\ldots)\nonumber\\
 &&+(e_{\nu+1-\r}+e_{\nu+3-\r}+\ldots),
\end{eqnarray}
\begin{eqnarray}
\label{bfup}
\hat{A}_{\r,\s,l}&=&e_{\nu-\s}\nonumber\\
\hat{u}_{\r,\s,l}&=&e_{\nu-\s}+e_{\r+N-1}+e_{N-l+1}\nonumber\\
\hat{Q}_{\r,\s,l}&=&(\s-1)\rho+(e_{\r+N-2}+e_{\r+N-4}
+\ldots)+(e_{N-l}+e_{N-l-2}+\ldots)\nonumber\\
 &&+(e_{\nu+1-\s}+e_{\nu+3-\s}+\ldots),
\end{eqnarray}
\begin{eqnarray}
\label{bfdown1}
\hat{A}_{\r,\s,l}&=&e_{\nu-\s}\nonumber\\
\hat{u}_{\r,\s,l}&=&e_{\nu-\s}+e_{\nu-\r}+e_{l-1}\nonumber\\
\hat{Q}_{\r,\s,l}&=&(e_{\nu-\s-1}+e_{\nu-\s-3}+\ldots)
 +(e_{\nu-\r-1}+e_{\nu-\r-3}+\ldots)\nonumber\\
 &&+(e_{l-2}+e_{l-4}+\ldots)
\end{eqnarray}
or
\begin{eqnarray}
\label{bfup1}
\hat{A}_{\r,\s,l}&=&e_{\s-1}\nonumber\\
\hat{u}_{\r,\s,l}&=&e_{\s-1}+e_{\r+N-1}+e_{N-l+1}\nonumber\\
\hat{Q}_{\r,\s,l}&=&(e_{\r+N-2}+e_{\r+N-4}+\ldots)
 +(e_{N-l}+e_{N-l-2}+\ldots)\nonumber\\
 &&+(e_{\s-2}+e_{\s-4}+\ldots)
\end{eqnarray}
where $\rho=e_1+e_2+\ldots+e_{\nu-2}$ and $e_a$ is the $\nu-2$
dimensional unit vector in the
$a$ direction, i.e
$(e_a)_b=\delta_{a,b}$ for $a\in \{1,2,\ldots,\nu-2\}$. We define
$e_a$ to be zero for $a\not\in\{
1,2,\ldots,\nu-2\}$.
The exponent $\eta$ is given in \cite{Date} and \cite{Date1} as
\begin{equation}
\eta=\frac{1}{4}(\r+l-\s-1)+\gamma(\r,l,\s)
\end{equation}
where $\gamma(\r,l,\s)=\frac{\r^2}{4(\nu+1-N)}+\frac{l^2}{4(N+2)}
-\frac{1}{8}-\frac{\s^2}{4(\nu+1)}$.
The relation between (\ref{bfdown}) and (\ref{bfup}) (and similarly
(\ref{bfdown1}) and (\ref{bfup1})) will become clear once we introduce
the polynomial expressions for the branching
functions in section \ref{sec-result} which correspond to the RSOS
configuration sums.

The polynomial generalizations of (\ref{bf}) for the
$\widehat{su}(2)_{\nu-N-1}\times \widehat{su}(2)_N/\widehat{su}(2)_{\nu-1}$
coset models are given in terms of
\begin{equation}
\label{X}
X_L^N[A,u,Q]=\sum_{m\geq 0,{\rm e.o.}\,{\rm restr.}\,Q}
q^{\frac{1}{4}mCm-\half Am} \prod_{i=1}^{\nu-2}
\left[ \begin{array}{c} \half (Im+u+Le_N)_i\\ m_i \end{array} \right]
\end{equation}
where the argument $Q\in \Ztwo$ and the arguments $A$ and
$u$ are $(\nu-2)$ dimensional vectors
with integer values.
We will actually find that the polynomial form for the branching
functions are given as linear combinations
of $X_L^N$ of the form
\begin{equation}
Y_L^N[A,u,Q;A',u',Q']=X_L^N[A,u,Q]-q^{\frac{L+1}{2}}X_L^N[A',u',Q']
\end{equation}
where sometimes the second term is absent (the detailed results are
presented in
section \ref{sec-result} in (\ref{result1}) and (\ref{result2})).
The branching functions of the coset conformal field theories can be obtained
from these polynomial expressions by taking the limit
$L\rightarrow \infty$ \cite{Kyoto}-\cite{Date1}.
Since we always assume $q<1$ the second term drops out in this limit but
this term will be essential for the proof of the
recursion relation. Notice also that the inhomogeneous term $L$ sits in
the $N^{\rm th}$ slot in (\ref{X}).
Using
$\lim_{n\rightarrow \infty} \left[ \begin{array}{c} n \\ m \end{array}
\right]=\frac{1}{(q)_m}$
one finds that in the limit
$L \rightarrow \infty$ the variable $m_N$ is treated differently from
the other $m$'s.
This agrees with the results in \cite{Kedem}
for $N=1$ and \cite{Gepner} for $N=2$.

The plan of the paper is as follows. In section \ref{sec-RSOS} we review the
general RSOS models and their relation to the conformal
coset models $\widehat{su}(2)_{\nu-N-1}\times \widehat{su}(2)_N/
\widehat{su}(2)_{\nu-1}$.
The path space interpretation and the recursion relation
for the configuration sums of the RSOS models will be essential
for the proof of the fermionic
forms given in this paper.
In section \ref{sec-result} we will state the explicit fermionic
polynomial expressions
of the coset models.
The proof of these formulas will be presented in section \ref{sec-proof}.
In subsection \ref{sec-tele} we will
get acquainted with the telescopic expansion technique and state
three important lemmas the proof of which is reserved
for the appendix. In subsection \ref{sec-initial}, \ref{sec-straight}
and \ref{sec-recursion} we prove the initial
conditions, the equality of two forms for the ``straight'' characters
$X(s|r,r)$ and the recursion relations respectively.
We conclude in section \ref{sec-duality} with a discussion of the
transformation of the polynomial forms under the duality
$q\rightarrow q^{-1}$ which correspond to the configuration sums
of the RSOS model in regime II and in the critical region
to the branching functions of the coset model
$\widehat{su}(2\nu-2)_1/\widehat{sp}(2\nu-2)_1$
\cite{Date}, \cite{Date1}. These branching functions can also
be identified with the $Z_{\nu-1}$-parafermion branching
functions.

\section{The general RSOS model}
\label{sec-RSOS}
\setcounter{equation}{0}

It was first observed by Huse \cite{Huse} that the configuration sum of the
ABF model \cite{ABF} in regime III are off-critical extensions of the branching
functions of the unitary minimal model $M(p,p+1)$. Since then many connections
between statistical mechanical configuration sums and conformal field theory
characters/branching functions have been revealed. For example, in the other
regimes the ABF model corresponds to the minimal models $M(2p-1,2p+1),
M(2,2p+1)$ and the $Z_p$ parafermion model. The ABF model has been generalized
by the Kyoto group \cite{Kyoto}-\cite{Date1} to a two parameter RSOS model.
This RSOS model in regime III corresponds to the
$\widehat{su}(2)_{\nu-N-1}\times \widehat{su}(2)_{N}/\widehat{su}(2)_{\nu-1}$
coset model and the RSOS model in regime II corresponds to the coset model
$\widehat{su}(2\nu-2)_1/\widehat{sp}(2\nu-2)_1$ and also to the $Z_{\nu-1}$
parafermion model. The configuration sum of the RSOS model in regime III and
regime II are related by the dual transformation $q\rightarrow q^{-1}$. We
will use this correspondence between the RSOS model and the conformal coset
models to prove the fermionic representation of the
$\widehat{su}(2)_{\nu-N-1}\times \widehat{su}(2)_{N}/\widehat{su}(2)_{\nu-1}$
branching functions.

The states of the general RSOS models can be described by paths and the
configuration sums are weighted sums over all possible paths
in the following way. Consider a square lattice
${\cal L}$. To each site $i$ of ${\cal L}$ one associates a state
variable $l_i$ which can take the values $l_i=1,2,\ldots,\nu$.
Two adjacent state variables $l_i$ and $l_{i+1}$ are called {\em admissable}
($l_i \sim l_{i+1}$) if they fulfill the following conditions:
\begin{eqnarray}
\label{admissable}
l_i-l_{i+1}=-N,-N+2,\ldots,N\nonumber \\
l_i+l_{i+1}=N+2,N+4,\ldots,2\nu-N
\end{eqnarray}
where $\nu$ and $N$ are arbitrary positive integers and $i$ runs
from $1,2,\ldots,L+2$. In terms of the path space $N$ defines the value for the
highest possible step in the vertical direction, $\nu$ and $L$
define the boundaries of the lattice in the vertical and horizontal
direction, respectively. The one dimensional configuration sum
for the RSOS models is given by \cite{ABF}, \cite{Date}, \cite{Date1}
\begin{equation}
X_{L}(a|b,c)=\sum_{l_i}q^{\sum_{j=1}^{L} \frac{j}{4} |l_{j+2}-l_j|}
\end{equation}
where $l_1=a, l_{L+1}=b, l_{L+2}=c$ and $l_1\sim l_2\sim \cdots \sim l_{L+2}$.
$X_L$ is uniquely determined by 1)~the initial condition
\begin{equation}
X_0(a|b,c)=\delta_{a,b}
\end{equation}
and 2) the recursion relation
\begin{equation}
\label{rec}
X_L(a|b,c)=\sum_{d\sim b}^{\: \: ''} q^{\frac{L}{4}|d-c|} X_{L-1}(a|d,b)
\end{equation}
where $\sum^{''}$ is the sum over d such that $d\sim b$ and $X_L$ is defined
to be zero if $a,b,c$ do not satisfy (\ref{admissable}).

The bosonic form for the configuration sum of the RSOS lattice models was
first calculated by ABF \cite{ABF}
for the case $N=1$ (in particular the Ising model and the hard square gas)
and for general $N$ by the Kyoto group \cite{Date},
\cite{Date1}. The corresponding characters for the conformal field theories
for $N=1$ were given by Rocha-Caridi \cite{RC}
based on work by Feigin and Fuchs \cite{FF}. The characters and branching
functions for $N=2$ have been given in \cite{GKO} and for the general
coset models
$\widehat{su}(2)_{\nu-N-1}\times \widehat{su}(2)_N/\widehat{su}(2)_{\nu-1}$
in \cite{KB}.

Because of the equivalence between the coset conformal field theories
$\widehat{su}(2)_{\nu-N-1}\times \widehat{su}(2)_N/\widehat{su}(2)_{\nu-1}$
and the
general RSOS models at the critical point one can interpret the expressions
for the branching function of a conformal field theory also as
the partition functions of the RSOS model in the limit $L\rightarrow \infty$.
The interpretation in terms of the RSOS models has the
advantage that the partition functions have finite dimensional polynomial
representations which obey the recursion relations (\ref{rec}).
The existence of these recursion relations enables us to prove the fermionic
representations of the partition functions.

We will prove the recursion relations in the spirit of \cite{Berkovich}
using telescopic
expansions. The telescopic expansions are based on the recursion
relations of the Gaussian polynomials
\begin{eqnarray}
\label{recursion'}
\left[ \begin{array}{c} n \\ m \end{array} \right] &=&
q^m \left[ \begin{array}{c} n-1 \\ m \end{array} \right] +
\left[ \begin{array}{c} n-1 \\ m-1 \end{array} \right] \\
\label{recursion}
\left[ \begin{array}{c} n \\ m \end{array} \right] &=&
\left[ \begin{array}{c} n-1 \\ m \end{array} \right] +
q^{n-m} \left[ \begin{array}{c} n-1 \\ m-1 \end{array} \right]
\end{eqnarray}
Notice that the recursion relations do not hold
for $m=n=0$. In this case one gets the contradiction $1=0$.
This leads to some complications for nonunitary models \cite{Will}
but won't be encountered in this paper.

\section{Fermionic form of the configuration sums of the RSOS model
and branching functions}
\label{sec-result}
\setcounter{equation}{0}

In this section we present the fermionic polynomial expression for
the configuration sums of the RSOS model in regime III
in the form of (\ref{X}). To this end let us define
\begin{eqnarray}
\label{down}
A^{{\rm down}}_{r,s}&=&e_{s-1}\nonumber\\
u^{{\rm down}}_{r,s}&=&e_{s-1}+e_{\nu-r}\nonumber\\
Q^{{\rm down}}_{r,s}&=&(r-1)\rho+(e_{s-2}+e_{s-4}+\ldots)
+(e_{\nu+1-r}+e_{\nu+3-r}+\ldots)\nonumber \\
	&&+L(e_{N-1}+e_{N-3}+\ldots)
\end{eqnarray}
and
\begin{eqnarray}
A^{{\rm up}}_{r,s}&=&e_{\nu-s}\nonumber\\
u^{{\rm up}}_{r,s}&=&e_{\nu-s}+e_{r-1}\nonumber\\
Q^{{\rm up}}_{r,s}&=&(s-1)\rho+(e_{r-2}+e_{r-4}+\ldots)
+(e_{\nu+1-s}+e_{\nu+3-s}+\ldots) \nonumber\\
	&&+L(e_{N-1}+e_{N-3}+\ldots)
\label{up}
\end{eqnarray}
where again $\rho=e_1+e_2+e_3+\ldots+e_{\nu-2}$ and $e_a$
is the $\nu-2$ dimensional unit vector in the $a$ direction, i.e
$(e_a)_b=\delta_{a,b}$ for $a\in \{1,2,\ldots,\nu-2\}$.
We define $e_a$ to be zero for $a\not\in\{1,2,\ldots,\nu-2\}$.

Our result for the fermionic form of the RSOS configuration
sum is given by
\begin{eqnarray}
\label{result1}
X_L(s|r,r+N)&\! \! =&\! \!
X_L^N[A_{r,s}^{{\rm down}},u_{r,s}^{{\rm down}},Q_{r,s}^{{\rm down}}]
\nonumber \\
X_L(s|r,r-N)&\! \! =&\! \!
X_L^N[A_{r,s}^{{\rm up}},u_{r,s}^{{\rm up}},Q_{r,s}^{{\rm up}}]
\end{eqnarray}
and
\begin{eqnarray}
\label{result2}
\lefteqn{X_L(s|r,r+N-2n)=}\nonumber\\
&&X_L^N[A_{r-n,s}^{{\rm down}},u_{r-n,s}^{{\rm down}}
+e_n,Q_{r-n,s}^{{\rm down}}
+(e_{n-1}+e_{n-3}+\ldots)]\nonumber \\
&&-\theta(r>n+1)q^{\frac{L+1}{2}}X_L^N[A_{r-n-1,s}^{{\rm down}},
 u_{r-n-1,s}^{{\rm down}}+e_{n-1},Q_{r-n-1,s}^{{\rm down}}+(e_{n-2}
+e_{n-4}+\ldots)]
\nonumber \\
&&\nonumber\\
\lefteqn{X_L(s|r,r-N+2n)=}\nonumber\\
&&X_L^N[A_{r+n,s}^{{\rm up}},u_{r+n,s}^{{\rm up}}+e_n,Q_{r+n,s}^{{\rm up}}
+(e_{n-1}+e_{n-3}+\ldots)]\nonumber \\
&&-\theta(r<\nu-n)q^{\frac{L+1}{2}}
X_L^N[A_{r+n+1,s}^{{\rm up}},u_{r+n+1,s}^{{\rm up}}
+e_{n-1},Q_{r+n+1,s}^{{\rm up}}
+(e_{n-2}+e_{n-4}+\ldots)]
\nonumber\\
\end{eqnarray}
for $n=1,2,\ldots,[\frac{N}{2}]$ where $[x]$ denotes the integer part of $x$
and $\theta(x>a)=\left\{ \begin{array}{ll} 1 & {\rm for} \; x>a\\
0 & {\rm otherwise} \end{array} \right. $.
$X_L(s|r,r+k)$ is zero if $s,r$ and $r+k$ are not admissable.

For $N=1$ the polynomials in (\ref{result1}) with $A,u,Q$ as given in
(\ref{down}) and (\ref{up}) coincide with Melzer's polynomials \cite{Melzer}
(the up and down cases in this paper are encoded by $L\not \equiv s-r \bmod 2$
and $L\equiv s-r \bmod 2$ in Melzer's paper). Melzer gives a second set of
fermionic expressions for the configuration sum. This set can also be
generalized to higher $N$
\begin{eqnarray}
\label{down1}
A^{{\rm down}}_{r,s}&=&e_{\nu-s}\nonumber\\
u^{{\rm down}}_{r,s}&=&e_{\nu-s}+e_{\nu-r}\nonumber\\
Q^{{\rm down}}_{r,s}&=&(e_{\nu-s-1}+e_{\nu-s-3}+\ldots)
+(e_{\nu-r-1}+e_{\nu-r-3}+\ldots)\nonumber \\
	&&+L(e_{N-1}+e_{N-3}+\ldots)
\end{eqnarray}
and
\begin{eqnarray}
A^{{\rm up}}_{r,s}&=&e_{s-1}\nonumber\\
u^{{\rm up}}_{r,s}&=&e_{s-1}+e_{r-1}\nonumber\\
Q^{{\rm up}}_{r,s}&=&(e_{r-2}+e_{r-4}+\ldots)
+(e_{s-2}+e_{s-4}+\ldots) \nonumber\\
	&&+L(e_{N-1}+e_{N-3}+\ldots)
\label{up1}
\end{eqnarray}
These two sets lead exactly to the same polynomials as the sets (\ref{down})
and (\ref{up}) and are merely a rewriting of the fermionic expressions. The
equivalence of both fermionic representations readily follows from lemma 1
(see below). Hence we will focus in the following on the set given in
(\ref{down}) and (\ref{up}).

Notice that there exists a symmetry between the ``up'' and the ``down''
expression for the configuration sum which
implies the mirror symmetry of the conformal grid. One may varify that
\begin{eqnarray}
A_{\tilde{r},\tilde{s}}^{{\rm up}}=A_{r,s}^{{\rm down}}\nonumber \\
u_{\tilde{r},\tilde{s}}^{{\rm up}}=u_{r,s}^{{\rm down}}\nonumber \\
Q_{\tilde{r},\tilde{s}}^{{\rm up}}=Q_{r,s}^{{\rm down}}
\end{eqnarray}
for both sets (\ref{down}), (\ref{up}) and (\ref{down1}), (\ref{up1}) where
$\tilde{r}=\nu+1-r$ and $\tilde{s}=\nu+1-s$. Hence it follows that
\begin{equation}
\label{symmetry}
X_L^{N-2n \: {\rm down}}(s|r,r+N-2n)=
X_L^{N-2n \: {\rm up}}(\tilde{s}|\tilde{r},\tilde{r}-(N-2n))
\end{equation}
for $n=0,1,2,\ldots,[\frac{N}{2}]$.

It is also important to notice that for even $N$ one gets two
different expressions for the ``straight'' character.
In this case $n=\frac{N}{2}$ and $X_L(s|r,r)$ is given once in
terms of $A^{{\rm up}}, u^{{\rm up}}, Q^{{\rm up}}$ and on the other hand
in terms of $A^{{\rm down}}, u^{{\rm down}}, Q^{{\rm down}}$.
We will prove the equality of both expressions in section \ref{sec-straight}.
Therefore in the proof of the recursion relations we can choose whichever
form is more convenient.

One obtains the branching functions of the rational coset conformal
field theories
\newline
$\widehat{su}(2)_{\nu-N-1}\times \widehat{su}(2)_N/\widehat{su}(2)_{\nu-1}$
by taking the limit $L\rightarrow \infty$ \cite{Date}, \cite{Date1}. Since we
always assume $q<1$ the second term in (\ref{result2}) drops out in the limit.
If one further uses
\begin{equation}
\lim_{n\rightarrow \infty} \left[ \begin{array}{c} n\\ m \end{array}\right] =
\frac{1}{(q)_m}
\end{equation}
one gets
\begin{equation}
\label{limit}
\lim_{\foot{$L\rightarrow \infty$}{$L$ \, even}} X_L(s|r,r+N-2n)=
\sum_{m\geq 0,{\rm e.o.}\,{\rm restr.}\,Q} q^{\frac{1}{4}mCm-\half Am}
\frac{1}{(q)_{m_N}}
 \prod_{\begin{tabular}{c} {\scriptsize $i=1$} \\
 {\scriptsize $i\neq N$} \end{tabular}}^{\nu-2}
 \left[ \begin{array}{c} \half(Im+u)_i \\ m_i \end{array} \right]
\end{equation}
for $n=0,1,\ldots,\left[ \frac{N}{2} \right]$ where from (\ref{down})
\begin{eqnarray}
A&=&e_{s-1}\nonumber\\
u&=&e_{s-1}+e_{\nu-r+n}+e_n\nonumber \\
Q&=&(r-n-1)\rho+(e_{s-2}+e_{s-4}+\ldots)+(e_{\nu+1-r+n}
+e_{\nu+3-r+n}+\ldots)\nonumber \\
 &&+(e_{n-1}+e_{n-3}+\ldots)
\end{eqnarray}
or from (\ref{down1})
\begin{eqnarray}
A&=&e_{\nu-s}\nonumber\\
u&=&e_{\nu-s}+e_{\nu-r+n}+e_n\nonumber \\
Q&=&(e_{\nu-s-1}+e_{\nu-s-3}+\ldots)+(e_{\nu-r+n-1}+e_{\nu-r+n-3}+\ldots)
\nonumber \\
 &&+(e_{n-1}+e_{n-3}+\ldots).
\end{eqnarray}

Similarly for $X_L(s|r,r-N+2n)$ one gets the same form as
in (\ref{limit}) for the
limit $L \rightarrow \infty$, $L$ even
but with
\begin{eqnarray}
A&=&e_{\nu-s}\nonumber \\
u&=&e_{\nu-s}+e_{r+n-1}+e_n\nonumber \\
Q&=&(s-1)\rho+(e_{r+n-2}+e_{r+n-4}+\ldots)+(e_{\nu+1-s}+e_{\nu+3-s}
+\ldots)\nonumber \\
 &&+(e_{n-1}+e_{n-3}+\ldots)
\end{eqnarray}
or
\begin{eqnarray}
A&=&e_{s-1}\nonumber \\
u&=&e_{s-1}+e_{r+n-1}+e_n\nonumber \\
Q&=&(e_{r+n-2}+e_{r+n-4}+\ldots)+(e_{s-2}+e_{s-4}+\ldots)\nonumber \\
 &&+(e_{n-1}+e_{n-3}+\ldots).
\end{eqnarray}

It has been established in \cite{Date} and \cite{Date1} that the limit of the
configuration sum of the RSOS
model equals the branching functions of the coset conformal field theory
\begin{equation}
q^{\eta}c_{\r\,\s}^{(l)}=\lim_{L\rightarrow \infty,L\,{\rm even}} X_L(a|b,c)
\end{equation}
where
\begin{eqnarray*}
&&\r=\half (b+c-N)\\
&&l=\half (b-c+N)+1\\
&&\s=a\\
&&\eta=\frac{1}{4}(b-a)-\gamma (\r,l,\s)
\end{eqnarray*}
and
\begin{equation}
\gamma (j_1,j_2,j_3)=\frac{j_1^2}{4m_1}
 +\frac{j_2^2}{4m_2}-\frac{1}{8}-\frac{j_3^2}{4m_3} \nonumber
\end{equation}
where in turn $m_1=\nu+1-N, m_2=N+2, m_3=\nu+1$.
Hence the polynomial forms (\ref{result1}) and (\ref{result2})
exactly lead to (\ref{bf}) with (\ref{bfdown}) or (\ref{bfdown1}) for the
``down'' case and (\ref{bfup}) or (\ref{bfup1}) for the ``up'' case in the
limit $L\rightarrow \infty, \,L$ even.

\section{The proof of the fermionic configuration sums}
\label{sec-proof}
\setcounter{equation}{0}

\subsection{The telescopic expansion technique}
\label{sec-tele}

In this section we set up the tools needed for the proof of
the fermionic sum formulas. We use the
telescopic expansion technique first developed in \cite{Berkovich}.
We introduce a shorthand notation (see (\ref{shorthand})),
and start by deriving important identities. At the end of this
subsection we state the general lemmas needed for the proof of
(\ref{result1})-(\ref{result2}). The proof of the lemmas is
reserved for the appendix.

Denote
\begin{equation}
\label{shorthand}
\bra{{\cal A}}{{\cal B}} \equiv
q^{\frac{1}{4}mCm-\half Am} \prod_{i=1}^{\nu-2}
\left[
\begin{array}{c}
\half (Im+u+Le_N+{\cal A})_i \\
(m+{\cal B})_i
\end{array}
\right]
\end{equation}
where $u$ and $A$ should be specified before each use of the abbreviation.
This notation is useful
since it avoids rewriting symbols that do not change during the calculations.
The entries ${\cal A}$ and ${\cal B}$ can arise
either from $u^{{\rm up}}$ and $u^{{\rm down}}$, variable changes
$m_i \rightarrow m_i+constant$ or the use of recursion relations
(\ref{recursion'}) and (\ref{recursion}).

For example for $\nu=8, N=2, s=1, r=5$, $L$ even and $A=u=0$ we have
\begin{eqnarray}
\label{ex}
X_L^2[A_{r,s}^{{\rm up}},u_{r,s}^{{\rm up}},Q_{r,s}^{{\rm up}}]&=&\! \!
\sum_{\foot{$ m_1, m_3$\, odd}{$ m_2, m_4, m_5, m_6$
 \, even}}
\left\{ \begin{array}{c} e_4 \\ 0 \end{array} \right\} \nonumber \\
&=&\sum_{m \, {\rm even}} q^{-\cir{1}-\cir{3}+1}
\left\{ \begin{array}{c} 2e_2+2e_4 \\ e_1+e_3 \end{array} \right\}
\end{eqnarray}
where we have made the variable changes
$m_1 \rightarrow m_1+1, \, m_3 \rightarrow m_3+1$
to obtain the second line. $\cir{i}$ is defined by
\begin{equation}
\cir{i}=\half (m_{i-1}+m_{i+1})-m_i.
\end{equation}
This combination always arises from the variable change in the
exponent $\frac{1}{4}mCm$ and also as the phase in the
recursion relation (\ref{recursion}).
We now perform our first telescopic expansion. Use the
recursion relation (\ref{recursion}) in the $4^{{\rm th}}$ slot
in (\ref{ex})
\begin{eqnarray}
\sum_{m \, {\rm even}} &&\left( q^{-\cir{1}-\cir{3}+1}
\left\{ \begin{array}{c} 2e_2 \\ e_1+e_3 \end{array} \right\} \right.
\nonumber \\
&&+ \left. q^{-\cir{1}-\cir{3}+\cir{4}+2}
\left\{ \begin{array}{c} 2e_2 \\ e_1+e_3-e_4 \end{array} \right\} \right)
\end{eqnarray}
Change variables $m_4 \rightarrow m_4+2$ in the second term
\begin{eqnarray}
\sum_{m \, {\rm even}} &&\left( q^{-\cir{1}-\cir{3}+1}
\left\{ \begin{array}{c} 2e_2 \\ e_1+e_3 \end{array} \right\}\right.
\nonumber \\
&&+ \left. q^{-\cir{1}-\cir{3}-\cir{4}+1}
\left\{ \begin{array}{c} 2e_2+2e_3+2e_5 \\ e_1+e_3+e_4 \end{array} \right\}
\right)
\end{eqnarray}
(Notice that the $2$ in the exponent becomes $1$ since one gets $2$ from
$\frac{1}{4}mCm$, $-2$ from $\cir{4}$ and $-1$ from $-\cir{3}$).
Repeating this procedure for the $5^{{\rm th}}$ slot etc. one finally obtains
\begin{eqnarray}
\label{ex1}
\sum_{m \, {\rm even}} &&\left( q^{-\cir{1}-\cir{3}+1}
\left\{ \begin{array}{c} 2e_2 \\ e_1+e_3 \end{array} \right\}\right.
\nonumber \\
&&+q^{-\cir{1}-\cir{3}-\cir{4}+1}
\left\{ \begin{array}{c} 2e_2+2e_3 \\ e_1+e_3+e_4 \end{array} \right\}
\nonumber \\
&&+ q^{-\cir{1}-\cir{3}-\cir{4}-\cir{5}+1}
\left\{ \begin{array}{c} 2e_2+2e_3+2e_4 \\ e_1+e_3+e_4+e_5 \end{array}
\right\}
\nonumber \\
&&+ \left. q^{-\cir{1}-\cir{3}-\cir{4}-\cir{5}-\cir{6}+1}
\left\{ \begin{array}{c} 2e_2+2e_3+2e_4+2e_5
\\ e_1+e_3+e_4+e_5+e_6 \end{array} \right\} \right)
\end{eqnarray}

In general there is an extra phase $q^{-\half m_{\nu-s}}$ or
$q^{-\half m_{s-1}}$ and an additional entry $e_{\nu-s}$
or $e_{s-1}$ in the ``up'' or ``down'' case respectively.
Then one gets an extra phase $\half$ when using the recursion relation
in the $(\nu-s)^{{\rm th}}$ or $(s-1)^{{\rm th}}$ slot which will
be turned into $-\half$ by doing the variable change $m_{\nu-s} \rightarrow
m_{\nu-s}+2$ or $m_{s-1} \rightarrow m_{s-1}+2$. Hence it is useful to define
\begin{equation}
\cir{i}'=\cir{i}+\half (A)_i.
\end{equation}

A general \underline{telescopic expansion} to the right of length
$M$ starting from $i$ with $A=u$ corresponding to recursion
relation (\ref{recursion}) is then given by
\begin{eqnarray}
\label{tele1}
&&\sum_{Q} q^{-\cir{i}'}
\left\{ \begin{array}{c} {\cal A}_{\leq i}+{\cal A}_{>i+M}+2e_{i+1}
\\ {\cal B}_{<i}+{\cal B}_{>i+M}+e_i \end{array} \right\} \nonumber \\
&=& \sum_{l=i}^{i+M} \,\sum_{Q}
q^{-\sum_{k=i}^{l}\cir{k}'} \left\{ \begin{array}{c} {\cal A}_{\leq i}
+{\cal A}_{>i+M}+2\sum_{k=i}^{l-1}e_k+2e_{i+M+1}\delta_{l,i+M} \\
{\cal B}_{<i}+{\cal B}_{>i+M}+\sum_{k=i}^{l}e_k \end{array} \right\}
\end{eqnarray}
for $i+M<\nu-2$.
Throughout the whole paper we define empty sums to be
zero (i.e. $\sum_{k=i}^{l}$ is zero if $l<i$) which avoids writing lots of
$\theta$-functions. Further $\sum_{Q}$ stands for
$\sum_{m\geq 0,{\rm e.o.}\,{\rm restr.}\, Q}$. This abbreviation will also
be used
in the rest of the paper. $Q\in\Ztwo$ is arbitrary and
${\cal A}_{\leq i}, {\cal A}_{>i+M}, {\cal B}_{<i}$ and ${\cal B}_{>i+M}$
are $\nu-2$ dimensional vectors with nonzero entries only in the slots
$\leq i, >i+M, <i$ and $>i+M$ respectively.
Notice that (\ref{tele1}) is also valid for $i+M=\nu-2$ that is the
telescopic expansion hits the end of the product of the Gaussians.
The term in the top row $2e_{i+M+1}\delta_{l,i+M}$ simply drops out.
(\ref{ex1}) is an example of this case.

One can derive (\ref{tele1}) in a similar fashion to (\ref{ex1})
by using repeatedly (\ref{recursion}) in the slots $k=i+1,i+2,\ldots,i+M$
followed by a variable change $m_k\rightarrow m_k+2$.
Another form of telescopic expansion corresponding to recursion relation
(\ref{recursion}) is given by
\begin{eqnarray}
\label{tele2}
&&\sum_{Q} q^{-\sum_{k=i}^{i+M}\cir{k}'}
\left\{ \begin{array}{c} {\cal A}_{\leq i}+{\cal A}_{>i+M}
+2\sum_{k=i}^{i+M}e_k
\\ {\cal B}_{<i}+{\cal B}_{>i+M}+\sum_{k=i}^{i+M}e_k \end{array}\right\}
\nonumber \\
&=&\sum_{l=i}^{i+M} \sum_{Q}
q^{-\sum_{k=i}^{l}\cir{k}'} \left\{ \begin{array}{c} {\cal A}_{\leq i}
+{\cal A}_{>i+M}+2\sum_{k=i}^{l-1}e_k
\\ {\cal B}_{<i}+{\cal B}_{>i+M}+\sum_{k=i}^{l}e_k \end{array}\right\}
\nonumber \\
&&+\sum_{Q}q^{\half (\A{\leq i})_i}
\left\{ \begin{array}{c} \A{\leq i} +\A{>i+M} \\ \B{<i}+\B{>i+M} \end{array}
\right\}
\end{eqnarray}
where we use a similar notation to (\ref{tele1}).
(\ref{tele2}) is derived via repeated application of (\ref{recursion})
starting from slot $i+M$. There are no variable changes involved.

The first type of telescopic expansion (\ref{tele1}) goes from left
to right whereas (\ref{tele2}) goes from right to left.
Analogous telescopic expansions of each type in the opposite direction
hold as well. There are also telescopic expansions corresponding
to recursion relation (\ref{recursion'}). All of these formulas are given
in appendix \ref{ap-identities}.

For the proof of (\ref{result1})-(\ref{result2}) we need further identities.
One of them is the \underline{mirror identity}
\begin{eqnarray}
\label{mirror}
&&\sum_{l=i}^{i+M} \, \sum_{Q}
q^{-\sum_{k=i}^{l}\cir{k}'} \left\{ \begin{array}{c} {\cal A}_{\leq i}
+{\cal A}_{>i+M}
+2\sum_{k=i}^{l-1}e_k \\ {\cal B}_{<i}+{\cal B}_{>i+M}+
\sum_{k=i}^{l}e_k \end{array} \right\}\nonumber \\
&=& \sum_{l=i}^{M+i} \, \sum_{Q}
q^{-\sum_{k=l}^{M+i}\cir{k}'+\frac{\theta(l>i)}{2}(\A{\leq i})_i}
\left\{ \begin{array}{c} {\cal A}_{\leq i}+{\cal A}_{>i+M}
+2\sum_{k=l+1}^{M+i}e_k
\\ {\cal B}_{<i}+{\cal B}_{>i+M}+\sum_{k=l}^{M+i}e_k \end{array} \right\}.
\end{eqnarray}
There again exists a mirror identity for recursion relation
(\ref{recursion'}) which is given in appendix~\ref{ap-identities}.

A mirror identity similar to (\ref{mirror}) has already been proven
in \cite{Berkovich}. Since we use a different notation in this paper
we repeat
the main steps of the proof. Define for $t=0,1,\ldots,M+1$
\begin{eqnarray}
\label{mirrorproof}
Z_t&=&\sum_{l'=i+M+1-t}^{i+M} \, \sum_{Q} q^{-\sum_{k=l'}^{i+M}\cir{k}'
+\frac{\theta(l'>i)}{2}(\A{\leq i})_i}
\left\{ \begin{array}{c} \A{\leq i}+\A{>i+M}+2\sum_{k=l'+1}^{M+i}e_k
\\ \B{<i}+\B{>i+M}+\sum_{k=l'}^{M+i}e_k \end{array} \right\} \nonumber \\
&+&\sum_{l=i}^{i+M-t}\sum_{Q} q^{-\sum_{k=i}^{l}\cir{k}'
-\sum_{k=i+M+1-t}^{i+M}\cir{k}'}\nonumber \\
&& \;\;\;\;\;\; \left\{ \begin{array}{c} \A{\leq i}+\A{>i+M}
+2\sum_{k=i}^{l-1}e_k+2\sum_{k=i+M+1-t}^{i+M}e_k
\\ \B{<i}+\B{>i+M}+\sum_{k=i}^{l}e_k +\sum_{k=i+M+1-t}^{i+M}e_k \end{array}
\right\}
\end{eqnarray}
where again empty sums are defined to be zero.
Notice that $Z_0=$lhs of (\ref{mirror}) and $Z_{M+1}=$rhs of (\ref{mirror}).
To prove that $Z_0=Z_1=\dots=Z_{M+1}$
expand the term with $l=i+M-t$ of the second sum in (\ref{mirrorproof})
via (\ref{tele2}) from $i+M-t-1$ to $i$
to the left. One then obtains
\begin{eqnarray}
\label{mirrorproof1}
&&\sum_{Q} q^{-\sum_{k=i}^{i+M}\cir{k}'}
\left\{ \begin{array}{c} \A{\leq i}+\A{>i+M}
+2\sum_{k=i}^{i+M-t-1}e_k+2\sum_{k=i+M+1-t}^{i+M}e_k
\\ \B{<i}+\B{>i+M}+\sum_{k=i}^{i+M}e_k \end{array} \right\}\nonumber \\
&=& \sum_{Q} \left( \sum_{l'=i}^{i+M-t-1} q^{-\sum_{k=i}^{l'}\cir{k}'
-\sum_{k=i+M-t}^{i+M}\cir{k}'}\right. \nonumber \\
&&\;\;\;\;\; \left\{ \begin{array}{c} \A{\leq i}+\A{>i+M}
+2\sum_{k=i}^{l'-1}e_k
+2\sum_{k=i+M+1-t}^{i+M}e_k
\\ \B{<i}+\B{>i+M}+\sum_{k=i}^{l'}e_k
+\sum_{k=i+M-t}^{i+M}e_k \end{array} \right\} \nonumber \\
&&+\left. q^{-\sum_{k=i+M-t}^{i+M}\cir{k}'+\half (\A{\leq i})_i}
\left\{ \begin{array}{c} \A{\leq i}+\A{>i+M}+2\sum_{k=i+M+1-t}^{i+M}e_k
\\ \B{<i}+\B{>i+M}+\sum_{k=i+M-t}^{i+M}e_k \end{array} \right\} \right).
\end{eqnarray}
The last term in (\ref{mirrorproof1}) is already the $l'=i+M-t$ term of the
first sum in $Z_{t+1}$. The first term in (\ref{mirrorproof1})
can be combined with the remaining terms in the second sum in
(\ref{mirrorproof}) in the $(i+M-t)^{\rm th}$ slot
\begin{eqnarray}
\lefteqn{\sum_{l=i}^{i+M-t-1}\sum_{Q}\!\!
q^{-\sum_{k=i}^{l}\cir{k}'-\sum_{i+M-t}^{i+M}\cir{k}'}}\nonumber \\
&&\bra{\A{\leq i}+\A{>i+M}+2\sum_{k=i}^{l-1}e_k
+2\sum_{k=i+M-t}^{i+M}e_k}{\B{<i}+\B{>i+M}+\sum_{k=i}^{l}e_k
+\sum_{k=i+M-t}^{i+M}e_k}
\end{eqnarray}
which is the second sum in $Z_{t+1}$. Hence (\ref{mirror}) is proven.

Further we will need the \underline{extended mirror identity}.
\begin{eqnarray}
\label{exmirror}
&&\sum_{Q}  \sum_{l=i+2}^{i+M}q^{-\cir{i}'-\sum_{k=l}^{i+M} \cir{k}'}
\bra{\A{\leq i}+\A{>i+M}+2e_{i+1}
+2\sum_{k=l+1}^{i+M}e_k}{\B{<i}+\B{>i+M}+e_i+\sum_{k=l}^{i+M}e_k } \nonumber \\
&=& \sum_{Q} \sum_{l=i}^{i+M-2} q^{-\cir{i+M}'-\sum_{k=i}^{l} \cir{k}'}
\bra{\A{\leq i}+\A{>i+M}+2e_{i+M-1}+2\sum_{k=i}^{l-1}e_k}{\B{<i}+\B{>i+M}
+e_{i+M}
+\sum_{k=i}^{l}e_k }
\end{eqnarray}
The proof of this formula goes as follows. Telescopically expand to different
lengths each term in the sum on the lhs via (\ref{tele1}) as follows.
Expand the $l^{\rm th}$ term to the right from $i$ to $l-2$. Hence there is
no expansion for the term $l=i+2$, however we expand the term
$l=i+3$ once, term $l=i+4$ twice etc. We then obtain
\begin{eqnarray}
\lefteqn{\sum_{Q} \, \sum_{l=i+2}^{i+M} \, \sum_{l'=i}^{l-2}\,
q^{-\sum_{k'=i}^{l'}\cir{k'}'-\sum_{k=l}^{i+M}\cir{k}'}}\nonumber \\
&& \bra{\A{\leq i}+\A{>i+M}+2\sum_{k'=i}^{l'-1}e_{k'}+2e_{l-1}
\delta_{l',l-2}+2\sum_{k=l+1}^{i+M}e_k}{\B{<i}+\B{>i+M}
+\sum_{k'=i}^{l'}e_{k'}+\sum_{k=l}^{i+M}e_k}
\end{eqnarray}
Exchanging the sums yields $\sum_{l'=i}^{i+M-2} \sum_{l=l'+2}^{i+M}$.
We can now recombine the $l$-sums via telescopic expansions
to the left and obtain the rhs of (\ref{exmirror}).

Finally, we need a \underline{special telescopic expansion}.
It will be needed when either $r$ or $s$ is even.
\begin{eqnarray}
\label{teleend}
\sum_{Q} \sum_{l=i}^{\nu-3} q^{-\sum_{k=i}^{l}\cir{k}'}
\bra{\A{\leq i}+2\sum_{k=i}^{l-1}e_k+e_{\nu-2}}{\B{<i}+\sum_{k=i}^{l}e_k}
\nonumber \\
=\sum_{Q} q^{-\cir{i}'-\sum_{k=i+2}^{\nu-2}\cir{k}'+\half}
\bra{\A{\leq i}+2\sum_{k=i+1}^{\nu-3}e_k+e_{\nu-2}}{\B{<i}+e_i
+\sum_{k=i+2}^{\nu-2}e_k}
\end{eqnarray}
The proof of this formula is similar to the one of the extended mirror
identity.
To this end we telescopically expand
each term on the lhs to the left starting from $\nu-2$. The $l^{\rm th}$ term
should be expanded to slot $l+2$
\begin{eqnarray}
\lefteqn{\sum_{Q} \, \sum_{l=i}^{\nu-3} \: \sum_{l'=l+2}^{\nu-1}
q^{-\sum_{k=i}^{l}\cir{k}'-\sum_{k'=l'}^{\nu-2}\cir{k'}'
+\half\theta(l'<\nu-1)}}\nonumber \\
&&\bra{\A{\leq i}+2\sum_{k=i}^{l-1}e_k+2\sum_{k'=l'+1}^{\nu-2}e_{k'}
+2e_{l+1}\delta_{l',l+2}-e_{\nu-2}}
{\B{<i}+\sum_{k=i}^{l}e_k+\sum_{k'=l'}^{\nu-2}e_{k'}}
\end{eqnarray}
Exchanging again the sums $\sum_{l'=i+2}^{\nu-1}\sum_{l=i}^{l'-2}$ and
recombining the $l$-sum via inverse telescopic expansion results in
\begin{eqnarray}
\sum_{Q}\, \sum_{l'=i+2}^{\nu-1} q^{-\sum_{k=l'}^{\nu-2}\cir{k}'
+\half\theta(l'<\nu-1)}
\bra{\A{\leq i}+2e_{i+1}+2\sum_{k'=l'+1}^{\nu-2}e_{k'}-e_{\nu-2}}{\B{<i}
+e_i+\sum_{k'=l'}^{\nu-2}e_{k'}}
\end{eqnarray}
Using then (\ref{tele2}) for the remaining sum in a slightly modified
version for slot $\nu-2$ yields the rhs of (\ref{teleend}).
The analogue of (\ref{teleend}) in opposite direction is given
in appendix \ref{ap-identities}.

With the help of the telescopic expansions and identities discussed
above we can prove three lemmas which
enable us to give the proof of the fermionic configuration sums.
The first lemma
makes it possible to transform the phase $q^{-\half m_{s-1}}$ and the
entry $e_{s-1}$ into $q^{-\half m_{\nu-s}}$ and $e_{\nu-s}$ respectively
and vice
versa. Since recursion relation (\ref{rec}) involves both $X^{{\rm up}}$ and
$X^{{\rm down}}$ we will be able to transform them into comparable formulas.
This lemma also establishes the equivalence between the two different sets
(\ref{down}), (\ref{up}) and (\ref{down1}), (\ref{up1}). Hence we can
restrict ourselves to the proof of (\ref{result1}) and (\ref{result2})
with $A,u,Q$ given by (\ref{down}) and (\ref{up}) in the following.

The configuration sums have the entry $\frac{L}{2}$ in the $N^{{\rm th}}$
slot. Since the telescopic expansions are sensitive to
the entries in the top line one has to treat the part to the right of the
$N^{{\rm th}}$ slot and the part to the left of the $N^{{\rm th}}$
slot separately. Lemmas 2 and 3 give the necessary tools to handle this
and will be frequently used in the following.

We now state the three lemmas:
\begin{lemma}
\label{lemma1}
With $A=u=0$ we have in the notation of (\ref{shorthand})
\begin{eqnarray}
\label{equ:lemma1}
\sum_{Q+(e_{s-2}+e_{s-4}+\ldots)} q^{-\half m_{s-1}}
\left\{
\begin{array}{c}
{\cal A} +e_{s-1} \\ 0
\end{array}
\right\} \nonumber \\
=\sum_{Q+(s-1)\rho +(e_{\nu+1-s}+e_{\nu+3-s}+\ldots)} q^{-\half m_{\nu-s}}
\left\{
\begin{array}{c}
{\cal A} +e_{\nu-s} \\ 0
\end{array}
\right\}
\end{eqnarray}
where ${\cal A}$ is an $\nu-2$ dimensional vector with integer values
and $Q \, \in ({\bf Z}_2)^{\nu-2}$ arbitrary.
\end{lemma}

Part a) and b) of the next lemma are useful for treating the slots to
the left of the $N^{{\rm th}}$ slot excluding and including the
$N^{{\rm th}}$ slot respectively for $N>2$.

\begin{lemma}
\label{lemma2}
{\bf a)} With $A=u$ we have in the notation of (\ref{shorthand})
\begin{eqnarray}
\label{equ:lemma2a}
&& \sum_{Q+(e_{a-1}+e_{a-3}+\ldots)}
\left\{ \begin{array}{c} e_{a}+{\cal B}_{\geq N}
\\ 0 \end{array} \right\} \nonumber \\
&=& \sum_{Q+(e_{\bar{a}}+e_{\bar{a}+2}+\ldots )_{<N}+a\rho_{<N}}
\left\{ \begin{array}{c} e_{\bar{a}-1}+e_{N-1}+{\cal B}_{\geq N}
\\ 0 \end{array} \right\} \nonumber \\
&-& \sum_{Q+(e_{\bar{a}-1}+e_{\bar{a}+1}+\ldots)_{<N}+a\rho_{<N}} q^{\half}
\left\{ \begin{array}{c} e_{\bar{a}-2}+e_{N}+{\cal B}_{\geq N}
\\ 0 \end{array} \right\}
\end{eqnarray}
where $Q\,\in \Ztwo$ arbitrary,
$\rho_{\footnotesize <N}=e_1+e_2+\ldots +e_{N-1}$, ${\cal B}_{\geq N}$
is a $(\nu-2)$ dimensional vector
that has non-negative values only for the entries bigger or equal
to $N$, i.e. $(\B{\geq N})_i=\left\{ \begin{array}{ll} 0
& {\rm for}\, 1\leq i<N\\ \in\, {\bf N}\cup \{0\} & {\rm for}\,
N\leq i\leq \nu-2
\end{array} \right.$, $a \, \in \{0,1,2,\ldots,N-2\}$
and $\bar{a}=N-a$. $(\ldots)_{<N}$ denotes
the projection onto the entries smaller than $N$, i.e.
$(e_n)_{<N}=\left\{ \begin{array}{ll} e_n & {\rm for} \, n<N \\
0 & {\rm for} \, n\geq N \end{array} \right. $.\\
{\bf b)} Under the same conditions as part a) we have
\begin{eqnarray}
\label{equ:lemma2b}
&& \sum_{Q+(e_{a}+e_{a-2}+\ldots)}
\left\{ \begin{array}{c} e_{a+1}+{\cal B}_{>N}
\\ 0 \end{array} \right\} \nonumber \\
&=& \sum_{Q+(e_{\bar{a}}+e_{\bar{a}+2}\ldots )_{<N}+(a-1)\rho_{<N}}
\left\{ \begin{array}{c} e_{\bar{a}-1}+e_{N}+{\cal B}_{>N} \\ 0
\end{array} \right\}
\nonumber \\
&-& \sum_{\foot{$Q+(a-1)\rho_{<N}+e_N$}{$+(e_{\bar{a}-1}
+e_{\bar{a}+1}+\ldots)_{<N}$}} q^{\frac{L+1}{2}}
\left\{ \begin{array}{c} e_{\bar{a}-2}+e_{N+1}+{\cal B}_{>N} \\ 0
\end{array} \right\}
\end{eqnarray}
where $\B{>N}$ is a $(\nu-2)$ dimensional vector that has
non-negative values only for the entries bigger than $N$.
\end{lemma}

The next lemma is the analogue of lemma (\ref{lemma2}) for the slots to the
right of the $N^{{\rm th}}$ slot.

\begin{lemma}
\label{lemma3}
{\bf a)} With $A=u$ we have in the notation of (\ref{shorthand})
\begin{eqnarray}
\label{equ:lemma3a}
&& \sum_{Q+(e_{r+a}+e_{r+a-2}+\ldots)_{\geq N}}
\left\{ \begin{array}{c} e_{r+a+1}+{\cal B}_{\leq N} \\ 0
\end{array} \right\} \nonumber \\
&=& \sum_{\foot{$Q+(r-\bar{a})\rho_{\geq N}+e_N$}{$+(e_{\nu-r+\bar{a}}
+e_{\nu-r+\bar{a}+2}+\ldots)$}}
\left\{ \begin{array}{c} e_{\nu-r+\bar{a}-1}+e_{N+1}+{\cal B}_{\leq N}
\\ 0 \end{array} \right\} \nonumber \\
&-& \sum_{\foot{$Q+(r-\bar{a}-1)\rho_{\geq N}$}{$+(e_{\nu-r+\bar{a}+1}
+e_{\nu-r+\bar{a}+3}+\ldots)$}} q^{\half}
\left\{ \begin{array}{c} e_{\nu-r+\bar{a}}+e_{N}+{\cal B}_{\leq N}
\\ 0 \end{array} \right\}
\end{eqnarray}
where $Q\, \in \Ztwo$ arbitrary, ${\cal B}_{\leq N}$ is a $(\nu-2)$
dimensional
vector that has non-negative values
only for the entries smaller or equal to $N$, $\rho_{\geq N}=e_N+e_{N+1}
+\ldots +e_{\nu-2}$,
$\bar{a}=N-a$ and $r+a>N$. $(\ldots)_{\geq N}$ denotes the projection
onto the entries
bigger or equal to $N$.\\
{\bf b)} Under the same condition as in part a) we have
\begin{eqnarray}
\label{equ:lemma3b}
&& \sum_{Q+(e_{r+a-1}+e_{r+a-3}+\ldots)_{\geq N}}
\left\{ \begin{array}{c} e_{r+a}+{\cal B}_{<N} \\ 0 \end{array} \right\}
\nonumber \\
&=& \sum_{\foot{$Q+(r-\bar{a})\rho_{\geq N}$}{$+(e_{\nu-r+\bar{a}}
+e_{\nu-r+\bar{a}+2}+\ldots)$}}
\left\{ \begin{array}{c} e_{\nu-r+\bar{a}-1}+e_{N}+{\cal B}_{<N}
\\ 0 \end{array} \right\} \nonumber \\
&-& \sum_{\foot{$Q+(r-\bar{a}-1)\rho_{\geq N}$}{$+(e_{\nu-r+\bar{a}+1}
+e_{\nu-r+\bar{a}+3}+\ldots)$}} q^{\frac{L+1}{2}}
\! \! \left\{ \begin{array}{c} e_{\nu-r+\bar{a}}+e_{N-1}+{\cal B}_{<N}
\\ 0 \end{array} \right\}
\end{eqnarray}
\end{lemma}
The proof of the three lemmas is given in appendix \ref{ap-lemmas}.

\subsection{Initial conditions}
\label{sec-initial}

In this section we prove the initial conditions for the configuration sums
\begin{eqnarray}
\label{initial}
X_0(s|r,r-N+2n)=\delta_{r,s} \nonumber \\
X_0(s|r,r+N-2n)=\delta_{r,s}
\end{eqnarray}
for $n=0,1,\ldots,[\frac{N}{2}]$.

First of all it suffices to show that the initial conditions hold for
$X_0^{\rm up}$ because by symmetry (\ref{symmetry})
it follows that
\begin{equation}
\label{pr-init}
X_0^{{\rm down}}(s|r,r+N-2n)=X_0^{{\rm up}}(\tilde{s}|\tilde{r},\tilde{r}-N+2n)
=\delta_{\tilde{r},\tilde{s}}=\delta_{r,s}.
\end{equation}
For $n=1,2,\ldots,[\frac{N}{2}]$ we have
\begin{eqnarray}
\label{pr-init1}
\lefteqn{X_0(s|r,r-N+2n)}\nonumber \\
&&=X_0^N[A_{r+n,s}^{{\rm up}},u_{r+n,s}^{{\rm up}}+e_n,Q_{r+n,s}^{{\rm up}}
+(e_{n-1}+e_{n-3}+\ldots)]\nonumber \\
&&-\theta(r<\nu-n)q^{\half}X_0^N[A_{r+n+1,s}^{{\rm up}},u_{r+n+1,s}^{{\rm up}}
+e_{n-1},Q_{r+n+1,s}^{{\rm up}}+(e_{n-2}+e_{n-4}+\ldots)]\nonumber \\
&&=\sum_{\foot{$Q+(e_{r+n-2}+e_{r+n-4}+\ldots)$}{$+(e_{n-1}+e_{n-3}+\ldots)$}}
\! \left\{ \begin{array}{c} e_{r+n-1}+e_n \\ 0 \end{array}\right\} \nonumber \\
&&-\theta(r<\nu-n)\sum_{\foot{$Q+(e_{r+n-1}+e_{r+n-3}+\ldots)$}{$+(e_{n-2}
+e_{n-4}+\ldots)$}} \!\! q^{\half}
\left\{ \begin{array}{c} e_{r+n}+e_{n-1} \\ 0 \end{array}\right\}
\end{eqnarray}
where we used the abbreviation (\ref{shorthand}) with $A=u=e_{\nu-s}$
and $Q=(s-1)\rho+(e_{\nu+1-s}+e_{\nu+3-s}+\ldots)$.
For $r=\nu-n$ (\ref{pr-init1}) equals to
\begin{eqnarray}
\sum_{\foot{$Q+(r-1)\rho$}{$+(e_{\nu+1-r}+e_{\nu+3-r}+\ldots)$}}
\bra{e_{\nu-r}}{0}
 =X_0^1[A_{r,s}^{\rm down},u_{r,s}^{\rm down},Q_{r,s}^{\rm down}]
\end{eqnarray}
(the equal sign follows by lemma \ref{lemma1} since it allows us to transform
the phase $q^{-\half m_{\nu-s}}$ into $q^{-\half m_{s-1}}$
and the entry $e_{\nu-s}$ into $e_{s-1}$).
For $2<r<\nu-n$ (\ref{pr-init1}) becomes
\begin{eqnarray}
\label{pr-init2}
\sum_{Q+(e_{r-2}+e_{r-4}+\ldots)}\bra{e_{r-1}}{0}
 =X_0^1[A_{r,s}^{\rm up},u_{r,s}^{\rm up},Q_{r,s}^{\rm up}]
\end{eqnarray}
To prove this let us first consider the case that $r$ is odd.
In this case (\ref{pr-init1}) becomes
\begin{eqnarray}
&&\sum_{Q+(e_{n+1}+e_{n+3}+\ldots+e_{r+n-2})}
\left\{ \begin{array}{c} e_{r+n-1}+e_n \\ 0 \end{array} \right\}\nonumber \\
&-&\sum_{Q+(e_{n}+e_{n+2}+\ldots+e_{r+n-1})} q^{\half}
\left\{ \begin{array}{c} e_{r+n}+e_{n-1} \\ 0 \end{array} \right\}.
\end{eqnarray}
Change variables $m_i\rightarrow m_i+1$ for $i=n+1,n+3,\ldots,r+n-2$
in the first
term and $i=n,n+2,\ldots,r+n-1$ in the
second term.
Performing then a telescopic expansion of type (\ref{tele1}) to the left and
using the mirror identity (\ref{mirror})
followed by the extended mirror identity (\ref{exmirror}) $\frac{r-3}{2}$
times for the first and second term we obtain
\begin{eqnarray}
\label{pr-init3}
\lefteqn{\sum_{l=r-2}^{r+n-2} \, \sum_{Q} q^{-\cir{1}'-\cir{3}'-\ldots
-\cir{r-4}'
-\sum_{k=r-2}^{l}\cir{k}'+\frac{r-1}{4}}}\nonumber \\
&&\;\;\;\;\bra{2e_2+2e_4+\ldots+2e_{r-3}
+2\sum_{k=r-2}^{l-1}e_k+2e_{r+n-1}}{e_1+e_3+\ldots+e_{r-4}+\sum_{k=r-2}^{l}e_k}
\nonumber \\
\lefteqn{-\sum_{l=r-2}^{r+n-3}\, \sum_{Q} q^{-\cir{1}'-\cir{3}'-\ldots
-\cir{r-4}'
-\sum_{k=r-2}^{l}\cir{k}'-\cir{r+n-1}'+\frac{r+1}{4}+\half}}\nonumber \\
&&\;\;\;\;\bra{2e_2+2e_4+\ldots+2e_{r-3}+2\sum_{k=r-2}^{l-1}e_k+2e_{r+n-2}
+2e_{r+n}}{e_1+e_3+\ldots+e_{r-4}+\sum_{k=r-2}^{l}e_k+e_{r+n-1}}
\end{eqnarray}
Changing variables $m_{r+n-1}\rightarrow m_{r+n-1}-2$ in the second
term we can combine the two sums in (\ref{pr-init3}) in
the $(r+n-1)^{{\rm th}}$ slot
\begin{eqnarray}
\lefteqn{\sum_{l=r-2}^{r+n-3}\, \sum_{Q} q^{-\cir{1}'-\cir{3}'
-\ldots-\cir{r-4}'-\sum_{k=r-2}^{l}\cir{k}'+\frac{r-1}{4}}}\nonumber \\
&&\;\;\; \bra{2e_2+2e_4+\ldots+2e_{r-3}+2\sum_{k=r-2}^{l-1}e_k}{e_1+e_3
+\ldots+e_{r-4}+\sum_{k=r-2}^{l}e_k} \nonumber \\
\lefteqn{+\sum_{Q}  q^{-\cir{1}'-\cir{3}'-\ldots-\cir{r-4}'
-\sum_{k=r-2}^{r+n-2}\cir{k}'+\frac{r-1}{4}}} \nonumber \\
&&\;\;\; \bra{2e_2+2e_4+\ldots+2e_{r-3}+2\sum_{k=r-2}^{r+n-3}e_k
+2e_{r+n-1}}{e_1+e_3+\ldots+e_{r-4}+\sum_{k=r-2}^{r+n-2}e_k}
\end{eqnarray}
Using now the telescopic expansion (\ref{tele1}) and changing
variables $m_i\rightarrow m_i-1$ for $i=1,3,\ldots,r-2$ gives
exactly (\ref{pr-init2}).

The proof of (\ref{pr-init2}) for $r$ even is very similar.
In this case one uses the telescopic expansion corresponding
to (\ref{teleend1}),
applies the mirror and extended mirror identites, changes variables
and combines the two sums again
in the $(r+n-1)^{{\rm th}}$ slot and uses (\ref{tele1}) to
obtain (\ref{pr-init2}).

The case $r=2$ is special and can only appear for $N=2, n=1$
assuming $0<n<\left[\frac{N}{2}\right]$ since otherwise $(r,r-N+2n)$
is not admissable.
For $N=2, n=1$ (\ref{pr-init1}) becomes
$X_0^1[A_{2,s}^{\rm down},u_{2,s}^{\rm down},Q_{2,s}^{\rm down}]$
the proof of which is left to the reader.

Hence we have shown that for $n=0,1,\ldots,\left[\frac{N}{2}\right]$
\begin{equation}
X_0^{{\rm up}}(s|r,r-N+2n)=\left\{ \begin{array}{ll}
X_0^1[A_{r,s}^{\rm down},u_{r,s}^{\rm down},Q_{r,s}^{\rm down}]
& {\rm for} \: r=\nu-n \: {\rm or}
 \: r=2\:{\rm and}\:n>1\\
X_0^1[A_{r,s}^{\rm up},u_{r,s}^{\rm up},Q_{r,s}^{\rm up}] &
 {\rm otherwise} \end{array} \right.
\end{equation}
These are exactly the $L=0$ ``up'' and ``down'' characters for $N=1$.
Hence the proof of (\ref{initial}) reduces
to the proof of
\begin{eqnarray}
\label{initialup}
X_0[A_{r,s}^{\rm up},u_{r,s}^{\rm up},Q_{r,s}^{\rm up}]=\delta_{r,s}\\
\label{initialdown}
X_0[A_{r,s}^{\rm down},u_{r,s}^{\rm down},Q_{r,s}^{\rm down}]=\delta_{r,s}.
\end{eqnarray}

Explicitly we have
\begin{eqnarray}
X_0[A_{r,s}^{\rm up},u_{r,s}^{\rm up},Q_{r,s}^{\rm up}]
=\!\!\!\!\!\sum_{\foot{$Q+(s-1)\rho+(e_{r-2}+e_{r-4}+\ldots)$}
 {$+(e_{\nu+1-s}+e_{\nu+3-s}+\ldots)$}}\!\!\!\!\!
q^{-\frac{m_{\nu-s}}{2}}\bra{e_{\nu-s}+e_{r-1}}{0}
\end{eqnarray}
where we used the abbreviation (\ref{shorthand}) with $A=u=0$.
$\bra{e_{\nu-s}+e_{r-1}}{0}$ is only nonzero if
there exists a solution for the $m$'s of the following set of inequalities
\begin{eqnarray}
&&0\leq m_1\leq \half (m_2+\delta_{r-1,1}+\delta_{\nu-s,1})
\nonumber \\
&&0\leq m_k\leq \half (m_{k-1}+m_{k+1}+\delta_{r-1,k}
+\delta_{\nu-s,k})\nonumber \\
&&0\leq m_{\nu-2}\leq \half (m_{\nu-3}+\delta_{r-1,\nu-2}
+\delta_{\nu-s,\nu-2}).
\end{eqnarray}
Adding all inequalities we get for $1<s<\nu$ and $1<r<\nu$ yields
\begin{equation}
m_1+m_{\nu-2}\leq 2
\end{equation}
For even $L$ (hence especially for $L=0$) $r-s$ is even.
It follows that $m_1$ is always odd and hence $m_1=1$.
Adding all but the first inequality yields $m_2+m_{\nu-2}\leq 3$.
We can further deduce from the first inequality that
$m_1=1\leq \half m_2$ and since $m_2$ is even it follows that
$m_2=2$. Proceeding this way for the case $r-1<\nu-s$
finally yields
\begin{eqnarray}
m_k=\left\{ \begin{array}{ll}k & {\rm for}\;k=1,2,\ldots,r-1\\
 r-1 & {\rm for}\;k=r,\ldots,\nu-s\\
 \nu+r-s-k-1 & {\rm for}\;k=\nu-s,\ldots,\nu-2 \end{array} \right.
\end{eqnarray}
The last inequality requires
$0\leq r-s+1\leq \half (r-s+2)$ from which follows that $r=s$.
One may varify that $\frac{1}{4}mCm=\frac{r-1}{2}$
for the above determined values for the $m$'s. Hence
$q^{\frac{1}{4}mCm-\half m_{\nu-s}}=q^0=1$ and we have proven
(\ref{initialup}) for $1<s<\nu$ and $1<r\leq \nu-s$.
All other cases can be proven in a very similar fashion.
Equation (\ref{initialdown}) follows again by the symmetry
argument (\ref{symmetry}).

\subsection{Proof of the equality of the two forms for the
``straight'' configuration sum}
\label{sec-straight}

In this section we want to show that the two forms for
the ``straight'' characters are equal. The ``straight'' characters
can only occur for $N=2n$ even and $n>0$ and are according to (\ref{result2})
given by
\begin{eqnarray}
\label{straightdown}
\lefteqn{X_L^{\rm (down)}(s|r,r)}\nonumber \\
&&=X_L^N[A_{r-n}^{\rm down},u_{r-n}^{\rm down}+e_n,Q_{r-n}^{\rm down}
+(e_{n-1}+e_{n-3}+\ldots)]\nonumber \\
&&-\theta(r>n+1)q^{\frac{L+1}{2}}
X_L^N[A_{r-n-1}^{\rm down},u_{r-n-1}^{\rm down}+e_{n-1},Q_{r-n-1}^{\rm down}
+(e_{n-2}+e_{n-4}+\ldots)]
\nonumber\\
\end{eqnarray}
and
\begin{eqnarray}
\label{straightup}
\lefteqn{X_L^{\rm (up)}(s|r,r)}\nonumber\\
&&=X_L^N[A_{r+n}^{\rm up},u_{r+n}^{\rm up}
+e_n,Q_{r+n}^{\rm up}+(e_{n-1}+e_{n-3}+\ldots)]\nonumber \\
&&-\theta(r<\nu-n)q^{\frac{L+1}{2}}X_L^N[A_{r+n+1}^{\rm up},u_{r+n+1}^{\rm up}
+e_{n-1},Q_{r+n+1}^{\rm up}+(e_{n-2}+e_{n-4}+\ldots)]
\nonumber\\
\end{eqnarray}
Using abbreviation (\ref{shorthand}) with $A=u=e_{s-1}$ and $Q=L(e_{N-1}
+e_{N-3}+\ldots)+(e_{s-2}+e_{s-4}+\ldots)$ and lemma 1
to convert the phase $q^{-\frac{m_{\nu-s}}{2}}$ and the entry $e_{\nu-s}$
into $q^{-\frac{m_{s-1}}{2}}$ and $e_{s-1}$ respectively
in (\ref{straightup}) we have explicitly
\begin{eqnarray}
\label{straightdowne}
\lefteqn{X_L^{\rm (down)}(s|r,r)}\nonumber \\
&&=\!\!\!\sum_{\foot{$Q+(r-n-1)\rho+(e_{n-1}+e_{n-3}
+\ldots)$}{$+(e_{\nu+1-r+n}+e_{\nu+3-r+n}+\ldots)$}}
 \!\!\!\bra{e_n+e_{\nu-r+n}}{0}\nonumber\\
&&-\theta(r>n+1)\!\!\!\!\!\!\!\! \sum_{\foot{$Q+(r-n)\rho
+(e_{n-2}+e_{n-4}+\ldots)$}{$+(e_{\nu+2-r+n}+e_{\nu+4-r+n}+\ldots)$}}
 \!\!\!\!\!\!\!\! q^{\frac{L+1}{2}}\bra{e_{n-1}+e_{\nu-r+n+1}}{0}
\end{eqnarray}
and
\begin{eqnarray}
\label{straightupe}
\lefteqn{X_L^{\rm (up)}(s|r,r)}\nonumber\\
&&=\sum_{\foot{$Q+(e_{n-1}+e_{n-3}+\ldots)$}{$+(e_{r+n-2}+e_{r+n-4}+\ldots)$}}
 \bra{e_n+e_{r+n-1}}{0}\nonumber\\
&&-\theta(r<\nu-n)\!\!\!\sum_{\foot{$Q+(e_{n-2}+e_{n-4}+\ldots)$}{$+(e_{r+n-1}
+e_{r+n-3}+\ldots)$}}
 \!\!\!q^{\frac{L+1}{2}}\bra{e_{n-1}+e_{r+n}}{0}.
\end{eqnarray}

Let us start with the easy case $r=n+1$. In this case
(\ref{straightdowne}) and (\ref{straightupe}) reduce to
\begin{equation}
\label{hit1}
X_L^{\rm (down)}(s|r,r)=\sum_{Q+(e_{n-1}+e_{n-3}+\ldots)} \bra{e_n}{0}
\end{equation}
and
\begin{eqnarray}
\label{hit}
X_L^{\rm (up)}(s|r,r)&=&\sum_{\foot{$Q
+(e_{N-1}+e_{N-3}+\ldots)$}{$+(e_{n-1}+e_{n-3}+\ldots)$}}
 \bra{e_n+e_N}{0}\nonumber\\
&-&q^{\frac{L+1}{2}}\sum_{\foot{$Q+(e_N+e_{N-2}
+\ldots)$}{$+(e_{n-2}+e_{n-4}+\ldots)$}}
 \bra{e_{n-1}+e_{N+1}}{0}
\end{eqnarray}
The restrictions in the first term of (\ref{hit}) can be
rewritten as $(e_{n+1}+e_{n+3}+\ldots)_{<N}+n\rho_{<N}$ and the one for the
second term as $(e_n+e_{n+2}+\ldots)_{<N}+n\rho_{<N}$.
The equality of (\ref{hit1}) and (\ref{hit}) hence follows by lemma 2b) with
$\bar{a}=n+1$ and $a=n-1$.

The equality of (\ref{straightupe}) and (\ref{straightdowne})
for $r=\nu-n$ follows by the symmetry (\ref{symmetry}) from the former case.

Let us now come to the proof for the case $n+1<r<\nu-n$. To this
end we split the first term
in (\ref{straightdowne}) via lemma 3b) with $\bar{a}=n+1$ and $a=N-n-1=n-1$.
\begin{eqnarray}
\label{straightdowne1}
\lefteqn{X_L^{\rm (down)}(s|r,r)}\nonumber \\
&&=\!\!\!\!\! \sum_{\foot{$Q+(r-n-1)\rho_{<N}+(e_{n-1}+e_{n-3}
+\ldots)$}{$+(e_{r+n-2}+e_{r+n-4}+\ldots)_{\geq N}$}}
 \!\!\!\bra{e_n+e_{r+n-1}-e_N}{0}\nonumber \\
&&+\!\!\!\!\!\sum_{\foot{$Q+(r-n-1)\rho_{<N}+(e_{n-1}+e_{n-3}
+\ldots)$}{$+(r-n)\rho_{\geq N}+(e_{\nu-r+n+2}+e_{\nu-r+n+4}+\ldots)$}}
 \!\!\!\!\! q^{\frac{L}{2}}\bra{e_n+e_{\nu-r+n+1}-e_N+e_{N-1}}{0}\nonumber \\
&&-\!\!\!\!\!\!\!\!\!\sum_{\foot{$Q+(r+N-1)\rho_{<N}+(e_{N-1}+e_{N-3}+\ldots)$}
 {$+(e_n+e_{n+2}+\ldots)_{<N}+(r-n)\rho_{\geq N}+(e_{\nu-r+n+2}+e_{\nu-r+n+4}
+\ldots)$}}
 \!\!\!\!\!\!\!\!\!\!\!\!\!\!\!\!\!\!\!\! q^{\frac{L+1}{2}}\bra{e_{n-1}
+e_{\nu-r+n+1}}{0}
\end{eqnarray}
Using lemma 2a) with $\bar{a}=n+1,a=n-1$ and ${\cal B}_{\geq N}
=-e_N+e_{\nu-r+n+1}$
to combine the second and third term in (\ref{straightdowne1})
we get
\begin{eqnarray}
\label{straightdowne2}
\lefteqn{X_L^{\rm (down)}(s|r,r)}\nonumber \\
&&\!\!\!\!\! =\sum_{\foot{$Q+(r-n-1)\rho_{<N}+(e_{n-1}+e_{n-3}
+\ldots)$}{$+(e_{r+n-2}+e_{r+n-4}+\ldots)_{\geq N}$}}
 \!\!\! \bra{e_n+e_{r+n-1}-e_N}{0}\nonumber \\
&&\!\!\!\!\! +\!\!\!\!\!\!\!\!\!\! \sum_{\foot{$Q+(r-n)\rho_{\geq N}
+(e_{\nu+2-r+n}+e_{\nu+4-r+n}+\ldots)$}{$+(r-n)\rho_{<N}+(e_{N-1}+e_{N-3}
+\ldots)
 +(e_{n-2}+e_{n-4}+\ldots)$}}
 \!\!\!\!\!\!\!\!\!\!\!\! q^{\frac{L}{2}}\bra{e_{n-1}+e_{\nu-r+n+1}-e_N}{0}
\end{eqnarray}
We can now combine the second term in (\ref{straightdowne2})
minus the second term in (\ref{straightupe}) where we use
$\tilde{r}=\nu+1-r$
\begin{eqnarray}
q^{\frac{L}{2}}&&\!\!\!\sum_{
 \begin{tabular}{c} {\scriptsize $Q+(r-n)\rho_{<N}+(e_{n-2}+e_{n-4}
+\ldots)$}\\
 {\scriptsize $+(e_{N-1}+e_{N-3}+\ldots)$}\\
 {\scriptsize $+(e_{\tilde{r}+n-1}+e_{\tilde{r}+n-3}+\ldots)_{\geq N}$}\\
 {\scriptsize $+(e_{\nu-2}+e_{\nu-4}+\ldots)_{\geq N}$}
 \end{tabular}}
 \bra{e_{n-1}+e_{\tilde{r}+n}-e_N}{0} \nonumber \\
&&\!\!\!\!\!\!\!+q^{\frac{L+1}{2}}\!\!\sum_{
 \begin{tabular}{c} {\scriptsize $Q+(r-n)\rho_{<N}+(e_{n-2}+e_{n-4}+\ldots)$}\\
 {\scriptsize $+(e_{N-1}+e_{N-3}+\ldots)$}\\
 {\scriptsize $+(\tilde{r}+n)\rho_{\geq N}+(e_{\nu-2}
+e_{\nu-4}+\ldots)_{\geq N}$}\\
 {\scriptsize $+(e_{\nu-\tilde{r}+n+2}+e_{\nu-\tilde{r}+n+4}+\ldots)$}
 \end{tabular}}
 \bra{e_{n-1}+e_{\nu+1-\tilde{r}+n}}{0} \nonumber \\
=q^{\frac{L}{2}}&&\!\!\!\!\!\!\!\!\!\!\!\!\!\! \sum_{
 \begin{tabular}{c} {\scriptsize $Q+(r-n)\rho_{<N}
+(e_{n-2}+e_{n-4}+\ldots)$}\\
 {\scriptsize $+e_N+(e_{N-1}+e_{N-3}+\ldots)$}\\
 {\scriptsize $+(e_{\nu-\tilde{r}+n+1}+e_{\nu-\tilde{r}+n+3}+\ldots)$}\\
 {\scriptsize $+(\tilde{r}-n-1)\rho_{\geq N}+(e_{\nu-2}
+e_{\nu-4}+\ldots)_{\geq N}$}
 \end{tabular}}
 \!\! \bra{e_{n-1}+e_{\nu+1-\tilde{r}+n}-e_N+e_N}{0}
\end{eqnarray}
where we obtained the last line by using lemma 3a)
with $a=n-1$ and $\bar{a}=n+1$. Hence we have
\begin{eqnarray}
\lefteqn{X_L^{\rm (down)}(s|r,r)-X_L^{\rm (up)}(s|r,r)}\nonumber \\
&=&\sum_{\foot{$Q+(r-n-1)\rho_{<N}+(e_{n-1}+e_{n-3}
+\ldots)$}{$+(e_{r+n-2}+e_{r+n-4}+\ldots)_{\geq N}$}}
 \bra{e_{n}+e_{r+n-1}-e_N}{0}\nonumber\\
&+&\sum_{\foot{$Q+(r-1)\rho_{<N}+(e_{n}+e_{n+2}+\ldots)$}{$+e_N
+(e_{r+n-2}+e_{r+n-4}+\ldots)_{\geq N}$}}
 q^{\frac{L}{2}}\bra{e_{N+1}+e_{r+n-1}-e_N+e_{n-1}}{0}\nonumber\\
&-&\sum_{\foot{$Q+(r-1)\rho_{<N}+(e_{n+1}+e_{n+3}+\ldots)_{<N}$}{$+(e_{r+n-2}
+e_{r+n-4}+\ldots)_{\geq N}$}}
 \bra{e_{r+n-1}+e_{n}}{0}
\end{eqnarray}
which is zero via lemma 2b) with $a=n-1$ and $\bar{a}=n+1$.

\subsection{Recursion relations}
\label{sec-recursion}

We now turn to the main part of the proof, the recursion relations.
We have to show that (\ref{result1})-(\ref{result2}) satisfy the
recursion relations (\ref{rec}) which reads for the
``down''-character
\begin{eqnarray}
\label{recdown}
X_L(s|r,r+N-2n)&=&\sum_{i=0}^{n}q^{(n-i)\frac{L}{2}}
X_{L-1}(s|r+N-2i,r)\nonumber\\
&&+\sum_{i=1}^{N-n}q^{i\frac{L}{2}}X_{L-1}(s|r+N-2n-2i,r)
\end{eqnarray}
where $n=0,1,2,\ldots,\left[\frac{N}{2}\right]$. Using the
symmetry (\ref{symmetry}) equation (\ref{recdown}) becomes
\begin{eqnarray}
\label{recup}
X_L(\tilde{s}|\tilde{r},\tilde{r}-N+2n)&=&\sum_{i=0}^{n}q^{(n-i)\frac{L}{2}}
X_{L-1}(\tilde{s}|\tilde{r}-N+2i,\tilde{r})\nonumber\\
&&+\sum_{i=1}^{N-n}q^{i\frac{L}{2}}
X_{L-1}(\tilde{s}|\tilde{r}-N+2n+2i,\tilde{r})
\end{eqnarray}
Hence the recursion relations for the ``up''-characters
follow automatically from the proof of (\ref{recdown})
for all admissable $r$ and $s$.
It is essential that we know the equality of the two
forms of $X(s|r,r)$ when using the symmetry argument here since
via (\ref{symmetry}) $X^{\rm down}(s|r,r)\rightarrow
X^{\rm up}(\tilde{s}|\tilde{r},\tilde{r})$. If $X^{\rm up}(s|r,r)
\neq X^{\rm down}(s|,r,r)$ (\ref{recup}) would not
follow from (\ref{recdown}).

Since (\ref{recup}) is implied by (\ref{recdown}) we can
restrict ourselves to the proof (\ref{recdown}). $r$ must fulfill
$\nu-N+n\geq r\geq n+1$ such that $(r,r+N-2n)$ is admissable.
Let us first consider the case that $\nu-N+n\geq r>
\left[ \frac{N+3}{2} \right]$ where $[x]$ denotes the integer part of $x$.
The characters $N$ ``up'' $X_{L-1}(s|r+N,r)$ to $N-2n$ ``up''
$X_{L-1}(s|r+N-2n,r)$ on the rhs of (\ref{recdown}) cancel pairwise
(the first term of $X_{L-1}(s|r+N-2i,r)$ with the second term of
$X_{L-1}(s|r+N-2(i+1))$) except for the first term of
$X_{L-1}(s|r+N-2n)$.
Similarly all ``down'' terms cancel ($X_{L-1}(s|r-N+2i,r)$ with
$i=0,1,\ldots,\left[\frac{N}{2}\right]$) except for the first
term of $X_{L-1}(s|r-N+2\left[\frac{N}{2}\right],r)$.
The characters
$X_{L-1}(s|r+N-2n-2,r)$ to $X_{L-1}(s|r+2,r)$ for
$N$ even or $X_{L-1}(s|r+1,r)$ for $N$ odd are all nonzero.
Hence (\ref{recdown}) becomes
\begin{eqnarray}
\label{recdownbig}
\lefteqn{X_L^N[A_{r-n,s}^{\rm down},u_{r-n,s}^{\rm down}
+e_n,Q_{r-n,s}^{\rm down}+(e_{n-1}+e_{n-3}+\ldots)]}\nonumber \\
\!&-&\!\!\! q^{\frac{L+1}{2}}\theta(n>0)
X_L^N[A_{r-n-1,s}^{\rm down},u_{r-n-1,s}^{\rm down}
+e_{n-1},Q_{r-n-1,s}^{\rm down}+(e_{n-2}+e_{n-4}+\ldots)]
\nonumber \\
\!&=&\!\!\! \sum_{k=n}^{\left[ \frac{N-1}{2} \right]} q^{(k-n)\frac{L}{2}}
X_{L-1}^{N}[A_{r+N-k,s}^{\rm up},u_{r+N-k,s}^{\rm up}+e_k,
Q_{r+N-k,s}^{\rm up}+(e_{k-1}+e_{k-3}+\ldots)]\nonumber \\
\!&-&\!\!\! \sum_{k=n}^{\left[ \frac{N-3}{2} \right]} q^{(k-n)\frac{L}{2}+L}
X_{L-1}^{N}[A_{r+N-k,s}^{\rm up},u_{r+N-k,s}^{\rm up}+e_k,
Q_{r+N-k,s}^{\rm up}+(e_{k-1}+e_{k-3}+\ldots)]\nonumber \\
\!&+&\!\!\!\! q^{\left( \left[ \frac{N+1}{2} \right]-n\right) \frac{L}{2}}
X_{L-1}^{N}[A_{r-\left[ \frac{N+1}{2} \right],s}^{\rm down},
u_{r-\left[ \frac{N+1}{2} \right],s}^{\rm down}
+e_{\left[ \frac{N}{2} \right]},
Q_{r-\left[ \frac{N+1}{2} \right],s}^{\rm down}
+e_{\left[ \frac{N}{2} \right]-1}
+e_{\left[ \frac{N}{2} \right]-3}+\ldots]\nonumber\\
\end{eqnarray}
for $n=0,1,2,\ldots,\left[\frac{N}{2}\right]$.

Further if $n+1<r\leq \left[ \frac{N+3}{2} \right]$
all ``down'' characters are zero since the endpoints are not admissable and
$X_{L-1}^{N-2k \, {\rm up}}(s|r+N-2k,r)$ is only nonzero
for $k\leq r-1$ since otherwise the pair $(r+N-2k,r)$ is not admissable.
Again the sum $\sum_{i=0}^{n}q^{(n-i)\frac{L}{2}}X_{L-1}(s|r+N-2i,r)$
cancels except for the first term
in $X_{L-1}(s|r+N-2n,r)$. Hence (\ref{recdown}) becomes
\begin{eqnarray}
\label{recdownsmall}
\lefteqn{X_L^N[A_{r-n,s}^{\rm down},u_{r-n,s}^{\rm down}
+e_n,Q_{r-n,s}^{\rm down}+(e_{n-1}+e_{n-3}+\ldots)]}\nonumber \\
&-&\!\!\!\! q^{\frac{L+1}{2}}\theta(n>0)
X_L^N[A_{r-n-1,s}^{\rm down},u_{r-n-1}^{\rm down}
+e_{n-1},Q_{r-n-1}^{\rm down}+(e_{n-2}+e_{n-4}+\ldots)]\nonumber \\
&=&\!\!\!\! \sum_{k=n}^{r-1}q^{(k-n)\frac{L}{2}}
X_{L-1}^N[A_{r+N-k,s}^{\rm up},u_{r+N-k,s}^{\rm up}
+e_k,Q_{r+N-k,s}^{\rm up}+(e_{k-1}+e_{k-3}+\ldots)]
\nonumber \\
&-&\!\!\!\! \sum_{k=n}^{r-2}q^{(k-n)\frac{L}{2}+L}
X_{L-1}^N[A_{r+N-k,s}^{\rm up},u_{r+N-k,s}^{\rm up}
+e_k,Q_{r+N-k,s}^{\rm up}+(e_{k-1}+e_{k-3}+\ldots)]
\nonumber\\
\end{eqnarray}
for $n=0,1,2,\ldots,\left[ \frac{N}{2}\right]$.

Finally for $r=n+1$ (\ref{recdown}) reduces to
\begin{eqnarray}
\label{recdowns}
&&X_L^N[A_{1,s}^{\rm down},u_{1,s}^{\rm down}
+e_n,Q_{1,s}^{\rm down}+(e_{n-1}+e_{n-3}+\ldots)]\nonumber \\
&=&X_{L-1}^N[A_{N+1,s}^{\rm up},u_{N+1,s}^{\rm up}
+e_n,Q_{N+1,s}^{\rm up}+(e_{n-1}+e_{n-3}+\ldots)].
\end{eqnarray}
Equation (\ref{recdownbig})-(\ref{recdowns}) are the
recursion relations we ought to prove.

We start with the proof of (\ref{recdownbig}). To this end we define
\begin{eqnarray}
\label{defF}
{\cal F}_b
=\sum_{k=n}^{\left[ \frac{N-1}{2} \right]-b} q^{(k-n)\frac{L}{2}}
X_{L-1}^{N}\!\!\!\!\!& [&\!\!\!\!\!
 A_{r+N-k,s}^{\rm up},u_{r+N-k,s}^{\rm up}+e_k,\nonumber \\
&&\!\!\!\!\! Q_{r+N-k,s}^{\rm up}+(e_{k-1}+e_{k-3}+\ldots)]\nonumber \\
-\sum_{k=n}^{\left[ \frac{N-3}{2} \right]-b} q^{(k-n)\frac{L}{2}+L}
X_{L-1}^{N}\!\!\!\!\! &[&\!\!\!\!\!
 A_{r+N-k,s}^{\rm up},u_{r+N-k,s}^{\rm up}+e_k,\nonumber \\
&&\!\!\! Q_{r+N-k,s}^{\rm up}+(e_{k-1}+e_{k-3}+\ldots)]\nonumber \\
+q^{\left( \left[ \frac{N+1}{2} \right]-n-b \right) \frac{L}{2}}
X_{L-1}^{N}\!\!\!\!\! &[&\!\!\!\!\!
 A_{r-\left[ \frac{N+1}{2} \right]+b,s}^{\rm down},
 u_{r-\left[ \frac{N+1}{2} \right]+b,s}^{\rm down}
+e_{\left[ \frac{N}{2} \right]+b},\nonumber \\
&&\!\!\!\!\! Q_{r-\left[ \frac{N+1}{2} \right]+b,s}^{\rm down}
+(e_{\left[ \frac{N}{2} \right]+b-1}+e_{\left[ \frac{N}{2} \right]+b-3}
+\ldots)]
\end{eqnarray}
for $b=0,1,\ldots,\left[ \frac{N-3}{2} \right]-n+1$.
For $b=\left[ \frac{N-3}{2} \right]-n+1$ the second sum is empty and
hence it drops out.
We are now going to show that
\begin{equation}
\label{Fequ}
{\cal F}_0={\cal F}_1=\cdots ={\cal F}_{\left[ \frac{N-3}{2}\right]-n+1}.
\end{equation}
Notice that ${\cal F}_0=$ the rhs of (\ref{recdownbig}).
Equation (\ref{Fequ}) follows if we can prove
\begin{eqnarray}
\label{claim}
X_{L-1}^N\!\!\!\!\!&[&\!\!\!\!\! A_{r+\left[ \frac{N}{2} \right]
+1+b}^{\rm up},
 u_{r+\left[ \frac{N}{2} \right]+1+b}^{\rm up}
+e_{\left[\frac{N-1}{2}\right]-b},
\nonumber \\
&&\!\!\!\!\! Q_{r+\left[ \frac{N}{2} \right]+1+b}^{\rm up}
 +(e_{\left[\frac{N-1}{2} \right]-b-1}
+e_{\left[\frac{N-1}{2}\right]-b-3}+\ldots)]\nonumber \\
-q^{\frac{L}{2}}X_{L-1}^N\!\!\!\!\!
&[&\!\!\!\!\! A_{r+\left[ \frac{N}{2} \right]+2+b}^{\rm up},
 u_{r+\left[ \frac{N}{2} \right]+2+b}^{\rm up}
+e_{\left[\frac{N-1}{2}\right]-1-b}, \nonumber\\
&&\!\!\!\!\! Q_{r+\left[ \frac{N}{2} \right]+2+b}^{\rm up}
 +(e_{\left[\frac{N-1}{2} \right]-b-2}
+e_{\left[\frac{N-1}{2}\right]-b-4}+\ldots)]\nonumber \\
+q^{\frac{L}{2}}X_{L-1}^N\!\!\!\!\! &[&\!\!\!\!\!
A_{r-\left[ \frac{N+1}{2} \right]+b}^{\rm down},
 u_{r-\left[ \frac{N+1}{2} \right]+b}^{\rm down}
+e_{\left[\frac{N}{2}\right]+b}, \nonumber\\
&&\!\!\!\!\! Q_{r-\left[ \frac{N+1}{2} \right]+b}^{\rm down}
 +(e_{\left[\frac{N}{2} \right]+b-1}
+e_{\left[\frac{N}{2}\right]+b-3}+\ldots)]\nonumber \\
=X_{L-1}^N\!\!\!\!\! &[&\!\!\!\!\!
A_{r-\left[ \frac{N+1}{2} \right]+(b+1)}^{\rm down},
 u_{r-\left[ \frac{N+1}{2} \right]+(b+1)}^{\rm down}
+e_{\left[\frac{N}{2}\right]+(b+1)}, \nonumber \\
&&\!\!\!\!\! Q_{r-\left[ \frac{N+1}{2} \right]
+(b+1)}^{\rm down}
 +(e_{\left[\frac{N}{2} \right]+b}+e_{\left[\frac{N}{2}\right]+b-2}+\ldots)]
\end{eqnarray}
where we dropped the subindex $s$. To prove this formula we define
$a=\left[ \frac{N}{2} \right]+b$, $\bar{a}=N-a=\left[\frac{N+1}{2} \right]
-b$, $Q=(e_{s-2}+e_{s-4}+\ldots)+(L-1)(e_{N-1}+e_{N-3}+\ldots)$ and
\begin{equation}
\label{newdef}
\bra{\cal A}{\cal B}\equiv q^{\frac{1}{4}mCm-\half m_{s-1}}\prod_{i=1}^{\nu-2}
\left[ \begin{array}{c} \half (Im+(L-1)e_N+e_{s-1}+{\cal A})_i \\
(m+{\cal B})_i \end{array} \right].
\end{equation}
(Notice that we changed $Le_N\rightarrow (L-1)e_N$ in the
definition which is more convenient since we deal with $X_{L-1}$).
Hence using lemma 1 on the first two terms in (\ref{claim})
to transform the phase $q^{\frac{m_{\nu-s}}{2}}$ into $q^{\frac{m_{s-1}}{2}}$
and the entry $e_{\nu-s}$ into $e_{s-1}$ the lhs of (\ref{claim}) becomes
\begin{eqnarray}
\label{claimexplicit}
&&\sum_{\foot{$Q+(e_{r+a-1}+e_{r+a-3}+\ldots)$}{$+(e_{\bar{a}-2}
+e_{\bar{a}-4}+\ldots)$}}\bra{e_{r+a}+e_{\bar{a}-1}}{0}\nonumber \\
&-&q^{\frac{L}{2}}\sum_{\foot{$Q+(e_{r+a}+e_{r+a-2}
+\ldots)$}{$+(e_{\bar{a}-3}+e_{\bar{a}-5}+\ldots)$}}
\bra{e_{r+a+1}+e_{\bar{a}-2}}{0}\nonumber \\
&+&q^{\frac{L}{2}}\sum_{\foot{$Q+(r-\bar{a}-1)\rho
+(e_{\nu+1-r+\bar{a}}$}{$+e_{\nu+3-r+\bar{a}}+\ldots)
+(e_{a-1}+e_{a-3}+\ldots)$}}
\bra{e_{\nu-r+\bar{a}}+e_{a}}{0}
\end{eqnarray}
Using lemma 2a) on the last term and splitting the
restrictions for the first two terms into
restrictions $\geq N$ and $<N$ (\ref{claimexplicit})
becomes
\begin{eqnarray}
&&\!\!\!\!\!\!\!\! \sum_{\foot{$Q+(r-N-1)\rho_{<N}
+(e_{\bar{a}}+e_{\bar{a}+2}+\ldots)_{<N}$}
{$+(e_{r+a-1}+e_{r+a-3}+\ldots)_{\geq N}$}} \bra{e_{r+a}
+e_{\bar{a}-1}}{0}\nonumber \\
-q^{\frac{L}{2}}&&\!\!\!\!\!\!\!\! \sum_{\foot{$Q+(r-N-1)
\rho_{<N}+(e_{\bar{a}-1}+e_{\bar{a}+1}+\ldots)_{<N}$}
{$+(e_{r+a}+e_{r+a-2}+\ldots)_{\geq N}$}} \bra{e_{r+a+1}
+e_{\bar{a}-2}}{0}\nonumber \\
+q^{\frac{L}{2}}&&\!\!\!\!\!\!\!\!
\sum_{\foot{$Q+(r-N-1)\rho_{<N}+(e_{\bar{a}}+e_{\bar{a}+2}+\ldots)_{<N}$}
{$+(r-\bar{a}-1)\rho_{\geq N}+(e_{\nu+1-r+\bar{a}}
+e_{\nu+3-r+\bar{a}}+\ldots)$}}
\!\! \bra{e_{\bar{a}-1}+e_{N-1}+e_{\nu-r+\bar{a}}}{0}\nonumber \\
-q^{\frac{L+1}{2}}&&\!\!\!\!\!\!\!\! \sum_{\foot{$Q
+(r-N-1)\rho_{<N}+(e_{\bar{a}-1}+e_{\bar{a}+1}+\ldots)_{<N}$}
{$+(r-\bar{a}-1)\rho_{\geq N}+(e_{\nu+1-r+\bar{a}}+e_{\nu+3-r
+\bar{a}}+\ldots)$}}
\!\!\! \bra{e_{\bar{a}-2}+e_{N}+e_{\nu-r+\bar{a}}}{0}
\end{eqnarray}
Using lemma 3b) to combine the first and third term
and lemma 3a) to combine the second and fourth term we get
(notice that we use the
lemmas with $L\rightarrow L-1$)
\begin{eqnarray}
&&\!\!\!\!\!\!\!\!\! \sum_{\foot{$Q+(r-N-1)\rho_{<N}
+(e_{\bar{a}}+e_{\bar{a}+2}+\ldots)_{<N}$}{$+(r-\bar{a})
\rho_{\geq N}+(e_{\nu-r+\bar{a}}
+e_{\nu-r+\bar{a}+2}+\ldots)$}}\!\!\!\!
\bra{e_{\bar{a}-1}+e_N+e_{\nu-r+\bar{a}-1}}{0}\nonumber \\
&-&\!\!\!\!\!\!q^{\frac{L}{2}}\!\!\!\!\!\!\! \sum_{\foot{$Q+(r-N-1)\rho_{<N}
+(e_{\bar{a}-1}+e_{\bar{a}+1}+\ldots)_{<N}$}{$+(r-\bar{a})\rho_{\geq N}
+e_N+(e_{\nu-r+\bar{a}}+e_{\nu-r+\bar{a}+2}+\ldots)$}}\!\!\!\!\!\!
\bra{\! e_{\bar{a}-2}+e_{N+1}+e_{\nu-r+\bar{a}-1}\! }{\!0\! }
\end{eqnarray}
which is via lemma 2b)
\begin{eqnarray}
\sum_{\foot{$Q+(r-\bar{a})\rho_{<N}
+(e_a+e_{a-2}+\ldots)$}{$(r-\bar{a})\rho_{\geq N}
 +(e_{\nu-r+\bar{a}}+e_{\nu-r+\bar{a}+2}+\ldots)$}}
\bra{e_{\nu-r+\bar{a}-1}+e_{a+1}}{0}\nonumber \\
=X_{L-1}^N[A_{r-(\bar{a}-1)}^{\rm down},u_{r-(\bar{a}-1)}^{\rm down}
+e_{a+1},Q_{r-(\bar{a}-1)}^{\rm down}+(e_{a}+e_{a-2}+\ldots)]
\end{eqnarray}
and hence (\ref{claim}) is proven.

With the help of (\ref{Fequ}) equation (\ref{recdownbig}) reduces to
\begin{eqnarray}
\label{recdownbig1}
&&\!\! X_L^N[A_{r-n}^{\rm down},u_{r-n}^{\rm down}
+e_n,Q_{r-n}^{\rm down}+(e_{n-1}+e_{n-3}+\ldots)]\nonumber \\
&&\!\!-q^{\frac{L+1}{2}}\theta(n>0)
X_L^N[A_{r-n-1}^{\rm down},u_{r-n-1}^{\rm down}+e_{n-1},
 Q_{r-n-1}^{\rm down}+(e_{n-2}+e_{n-4}+\ldots)]\nonumber \\
&=&\!\!\! X_{L-1}^N[A_{r+N-n}^{\rm up},u_{r+N-n}^{\rm up}
+e_n,Q_{r+N-n}^{\rm up}+(e_{n-1}+e_{n-3}+\ldots)]\nonumber \\
&&\!\!\!+q^{\frac{L}{2}}
X_{L-1}^N[A_{r-n-1}^{\rm down},u_{r-n-1}^{\rm down}
+e_{N-n-1},Q_{r-n-1}^{\rm down}+(e_{N-n-2}+e_{N-n-4}+\ldots)]
\end{eqnarray}
Notice that (\ref{recdownbig1}) immediately follows from (\ref{recdownbig})
for $N=1,2$ without any derivation. Hence (\ref{recdownbig1})
can be thought of as a generaliztion of the $N=1,2$ recursion relations.
Let us begin by showing that (\ref{recdownbig1}) holds for $n=0$. Using
lemma \ref{lemma1} again as before
(\ref{recdownbig1}) reads explicitly for $n=0$
\begin{eqnarray}
\label{recdownbig2}
&&\sum_{\foot{$Q+(r-1)\rho+(e_{N-1}+e_{N-3}+\ldots)$}{$+(e_{\nu+1-r}
+e_{\nu+3-r}+\ldots)$}}
\bra{e_{\nu-r}+e_N}{0}\nonumber \\
&=&\sum_{Q+(e_{r+N-2}+e_{r+N-4}+\ldots)}\bra{e_{r+N-1}}{0}\nonumber \\
&+&q^{\frac{L}{2}}\sum_{\foot{$Q+(r-2)\rho+(e_{N-2}+e_{N-4}
+\ldots)$}{$+(e_{\nu+2-r}+e_{\nu+4-r}+\ldots)$}}
\bra{e_{\nu-r+1}+e_{N-1}}{0}
\end{eqnarray}
where $Q$ and $\bra{{\cal A}}{{\cal B}}$ are defined as
in (\ref{newdef}). But (\ref{recdownbig2}) holds via lemma 3b) with $a=N-1$.
It remains to show that (\ref{recdownbig1}) is true for $n>0$.
In this case we have explicitly
\begin{eqnarray}
\label{recdownbig3}
&&\sum_{\foot{$Q+(r-n-1)\rho+(e_{\nu-r+n+1}+e_{\nu-r+n+3}
+\ldots)$}{$+(e_{N-1}+e_{N-3}+\ldots)+(e_{n-1}+e_{n-3}+\ldots)$}}
 \bra{e_{\nu-r+n}+e_N+e_n}{0}\nonumber\\
&-&\sum_{\foot{$Q+(r-n)\rho+(e_{\nu-r+n+2}+e_{\nu-r+n+4}
+\ldots)$}{$+(e_{N-1}+e_{N-3}+\ldots)+(e_{n-2}+e_{n-4}+\ldots)$}}
 q^{\frac{L+1}{2}}\bra{e_{\nu-r+n+1}+e_{N}+e_{n-1}}{0}\nonumber\\
&=&\sum_{\foot{$Q+(e_{n-1}+e_{n-3}+\ldots)$}{$+(e_{r+N-n-2}
+e_{r+N-n-4}+\ldots)$}}
\bra{e_n+e_{r+N-n-1}}{0}\nonumber\\
&+&\sum_{\foot{$Q+(r-n)\rho+(e_{N-n-2}+e_{N-n-4}
+\ldots)$}{$+(e_{\nu-r+n+2}+e_{\nu-r+n+4}+\ldots)$}}
q^{\frac{L}{2}}\bra{e_{\nu-r+n+1}+e_{N-n-1}}{0}
\end{eqnarray}
We can combine the second term on the lhs with the
second term on the rhs via lemma 2a) with $a=N-n-1$ and rewrite the
restrictions.
We then obtain for (\ref{recdownbig3})
\begin{eqnarray}
&&\!\!\!\sum_{\foot{$Q+(r-n-1)\rho_{\geq N}+(e_{\nu-r+n+1}
+e_{\nu-r+n+3}+\ldots)$}{$+(N+r-1)\rho_{<N}+(e_{n+1}+e_{n+3}+\ldots)_{<N}$}}
\!\!\! \bra{e_{\nu-r+n}+e_N+e_n}{0}\nonumber\\
&=&\!\!\! \sum_{\foot{$Q+(e_{n-1}+e_{n-3}
+\ldots)$}{$+(e_{r+N-n-2}+e_{r+N-n-4}+\ldots)$}}
\!\! \bra{e_n+e_{r+N-n-1}}{0}\nonumber\\
&+&\!\!\!\!\!\!\!\!\sum_{\foot{$Q+(r-n)\rho
+(e_{\nu-r+n+2}+e_{\nu-r+n+4}+\ldots)$}{$+(N-n-1)\rho_{<N}
+(e_{n+1}+e_{n+3}+\ldots)_{<N}$}}
\!\!\!\!\! q^{\frac{L}{2}}\bra{e_{\nu-r+n+1}+e_{N-1}+e_n}{0}
\end{eqnarray}
which is again true due to lemma 3b) with $a=N-n-1$.
Hence (\ref{recdownbig}) is proven.

We now turn to the proof of recursion relation (\ref{recdownsmall}).
Again we use the notation $\bra{{\cal A}}{{\cal B}}$ and the
definition of $Q$ as in (\ref{newdef}) and apply lemma \ref{lemma1}
to write (\ref{recdownsmall}) explicitly as
\begin{eqnarray}
\label{recdownsmall1}
&&\!\!\! \sum_{\foot{$Q+(r-n-1)\rho+(e_{\nu-r+n+1}+e_{\nu-r+n+3}+\ldots)$}
 {$+(e_{N-1}+e_{N-3}+\ldots)+(e_{n-1}+e_{n-3}+\ldots)$}}
 \bra{e_{\nu-r+n}+e_n+e_N}{0}\nonumber\\
&&\!\!\!-\theta(n>0)\!\!\!\!\!\!\!\!\!\!\!
 \sum_{\foot{$Q+(r-n)\rho+(e_{\nu-r+n+2}+e_{\nu-r+n+4}
+\ldots)$}{$+(e_{N-1}+e_{N-3}+\ldots)+(e_{n-2}+e_{n-4}+\ldots)$}}
 \!\!\!\!\!\!\! q^{\frac{L+1}{2}}\bra{e_{\nu-r+n+1}+e_{n-1}+e_N}{0}\nonumber \\
&=&\!\!\!\sum_{k=n}^{r-1}q^{(k-n)\frac{L}{2}}\sum_{\foot{$Q+(e_{k-1}
+e_{k-3}+\ldots)$}{$+(e_{r+N-k-2}+e_{r+N-k-4}+\ldots)$}}
 \bra{e_k+e_{r+N-k-1}}{0}\nonumber \\
&&\!\!\! -\sum_{k=n}^{r-2}q^{(k-n)\frac{L}{2}+L}\sum_{\foot{$Q+(e_{k-1}
+e_{k-3}+\ldots)$}{$+(e_{r+N-k-2}+e_{r+N-k-4}+\ldots)$}}
 \bra{e_k+e_{r+N-k-1}}{0}
\end{eqnarray}
We can combine the first term on the lhs with the term $k=n$ of the
first sum on the rhs via lemma 3b) with $a=N-n-1$
\begin{eqnarray}
&&\!\!\! \sum_{\foot{$Q+(r-N-1)\rho_{<N}+(e_{n+1}+e_{n+3}+\ldots)_{<N}$}
 {$(r-n-1)\rho_{\geq N}+(e_{\nu-r+n+1}+e_{\nu-r+n+3}+\ldots)$}}
 \!\!\bra{e_{\nu-r+n}+e_n+e_N}{0}\nonumber\\
&-&\!\!\!\sum_{\foot{$Q+(r-N-1)\rho_{<N}+(e_{n+1}+e_{n+3}
+\ldots)_{<N}$}{$+(e_{r+N-n-2}+e_{r+N-n-4}+\ldots)_{\geq N}$}}
 \!\!\bra{e_n+e_{r+N-n-1}}{0}\nonumber\\
&=&\!\!\!\!\!\! q^{\frac{L}{2}}\!\!\!\!\!\!
\sum_{\foot{$Q+(r-N-1)\rho_{<N}+(e_{n+1}+e_{n+3}+\ldots)_{<N}$}
 {$+(r-n)\rho_{\geq N}+(e_{\nu+2-r+n}+e_{\nu+4-r+n}+\ldots)$}}
 \!\!\!\bra{e_n+e_{N-1}+e_{\nu-r+n+1}}{0}
\end{eqnarray}
In the case that $n>0$ we can further combine this
result with the second term on the lhs of (\ref{recdownsmall1})
via lemma 2a) with $a=N-n-1$ and obtain
\begin{equation}
q^{\frac{L}{2}}\sum_{\foot{$Q+(r-n)\rho+(e_{N-n-2}
+e_{N-n-4}+\ldots)$}{$+(e_{\nu-r+n+2}+e_{\nu-r+n+4}+\ldots)$}}
 \bra{e_{N-n-1}+e_{\nu-r+n+1}}{0}
\end{equation}
Hence with the definition
\begin{eqnarray}
{\cal K}_n&\equiv &
 \sum_{\foot{$Q+(r-n)\rho+(e_{N-n-2}+e_{N-n-4}
+\ldots)$}{$+(e_{\nu-r+n+2}+e_{\nu-r+n+4}+\ldots)$}}
 \bra{e_{N-n-1}+e_{\nu-r+n+1}}{0}\nonumber\\
&-&\sum_{k=n}^{r-2}q^{(k-n)\frac{L}{2}}
\left( \sum_{\foot{$Q+(e_k+e_{k-2}+\ldots)$}
 {$+(e_{r+N-k-3}+e_{r+N-k-5}+\ldots)$}}
 \bra{e_{k+1}+e_{r+N-k-2}}{0} \right. \nonumber\\
&& \;\;\;\;\;\;\;\;\;\;
 -\left. q^{\frac{L}{2}}\sum_{\foot{$Q+(e_{k-1}
+e_{k-3}+\ldots)$}{$+(e_{r+N-k-2}+e_{r+N-k-4}+\ldots)$}}
 \bra{e_k+e_{r+N-k-1}}{0}\right)
\end{eqnarray}
for $n=0,1,\ldots,r-2$ we need to prove that ${\cal K}_n=0$.
First we show that similar to (\ref{Fequ})
\begin{equation}
{\cal K}_0={\cal K}_1=\cdots={\cal K}_{r-2}.
\end{equation}
To this end we need to derive for $b=0,1,2,\ldots,r-3$
\begin{eqnarray}
\label{claim1}
&&\sum_{\foot{$Q+(r-b)\rho+(e_{N-b-2}+e_{N-b-4}
+\ldots)$}{$+(e_{\nu+2+b-r}+e_{\nu+4+b-r}+\ldots)$}}
 \bra{e_{N-b-1}+e_{\nu-r+b+1}}{0}\nonumber\\
&-&\sum_{\foot{$Q+(e_b+e_{b-2}+\ldots)$}{$+(e_{r+N-b-3}
+e_{r+N-b-5}+\ldots)$}}
 \bra{e_{b+1}+e_{r+N-b-2}}{0}\nonumber\\
&+&\sum_{\foot{$Q+(e_{b-1}+e_{b-3}+\ldots)$}{$+(e_{r+N-b-2}
+e_{r+N-b-4}+\ldots)$}}
 q^{\frac{L}{2}}\bra{e_b+e_{r+N-b-1}}{0}\nonumber\\
&=&\!\!\! \sum_{\foot{$Q+(r+b+1)\rho+(e_{N-b-3}+e_{N-b-5}
+\ldots)$}{$+(e_{\nu+3+b-r}+e_{\nu+5+b-r}+\ldots)$}}
 \bra{e_{N-b-2}+e_{\nu-r+b+2}}{0}
\end{eqnarray}
We can apply lemma 2b) on the first term on the lhs with
$a=N-b-2$ and rewrite the restriction for the second and third term and get
\begin{eqnarray}
\lefteqn{{\rm lhs\, of\, (\ref{claim1})}}\nonumber \\
&=&\!\!\! \sum_{\foot{$Q+(N+r-1)\rho_{<N}+(e_{b+2}+e_{b+4}
+\ldots)_{<N}$}{$+(r+b)\rho_{\geq N}+(e_{\nu+2+b-r}+e_{\nu+4+b-r}+\ldots)$}}
 \!\!\! \bra{e_{\nu-r+1+b}+e_{b+1}+e_N}{0}\nonumber\\
&-&\!\!\!\!\!\!\! \sum_{\foot{$Q+(N+r-1)\rho_{<N}+e_N
+(e_{b+1}+e_{b+3}+\ldots)_{<N}$}
 {$+(r+b)\rho_{\geq N}+(e_{\nu+2+b-r}+e_{\nu+4+b-r}+\ldots)$}}
 \!\!\! q^{\frac{L}{2}}\bra{e_{\nu-r+1+b}+e_b+e_{N+1}}{0}\nonumber\\
&-&\!\!\!\!\!\! \sum_{\foot{$Q+(r+N-1)\rho_{<N}+(e_{b+2}
+e_{b+4}+\ldots)_{<N}$}{$+(e_{r+N-b-3}+e_{r+N-b-5}+\ldots)_{\geq N}$}}
 \!\!\! \bra{e_{b+1}+e_{r+N-b-2}}{0}\nonumber\\
&+&\!\!\!\!\!\! \sum_{\foot{$Q+(r+N-1)\rho_{<N}+(e_{b+1}
+e_{b+3}+\ldots)_{<N}$}{$+(e_{r+N-b-2}+e_{r+N-b-4}+\ldots)_{\geq N}$}}
 \!\!\! q^{\frac{L}{2}}\bra{e_b+e_{r+N-b-1}}{0}\nonumber\\
\end{eqnarray}
Combining now the first and third term via lemma 3b) and the
second and fourth term via lemma 3a) with $a=N-b-2$ in both cases we get
\begin{eqnarray}
\lefteqn{{\rm lhs\, of\, (\ref{claim1})}}\nonumber \\
&=&\!\!\!\!\! \sum_{\foot{$Q+(r+N-1)\rho_{<N}+(e_{b+2}+e_{b+4}+\ldots)_{<N}$}
 {$+(r-b-1)\rho_{\geq N}+(e_{\nu-r+b+3}+e_{\nu-r+b+5}+\ldots)$}}
 \!\!\!\!\!q^{\frac{L}{2}}\bra{e_{b+1}+e_{N-1}+e_{\nu-r+b+2}}{0}\nonumber \\
&-&\!\!\!\!\! \sum_{\foot{$Q+(r+N-1)\rho_{<N}+(e_{b+1}+e_{b+3}+\ldots)_{<N}$}
 {$+(r-b-1)\rho_{\geq N}+(e_{\nu-r+b+3}+e_{\nu-r+b+5}+\ldots)$}}
 \!\!\!\!\! q^{\frac{L+1}{2}}\bra{e_b+e_N+e_{\nu-r+b+2}}{0}
\end{eqnarray}
Via lemma 2a) with $a=N-b-2$ this is equal to the rhs of (\ref{claim1}).

Hence it remains to show that ${\cal K}_{r-2}=0$. This equation reads
explicitly
\begin{eqnarray}
&&\!\!\!\!\!\sum_{Q+(e_{N-r}+e_{N-r-2}+\ldots)} \! \bra{e_{N-r+1}}{0}
\nonumber\\
&-&\!\!\! \sum_{\foot{$Q+(e_{N-1}+e_{N-3}+\ldots)$}{$+(e_{r-2}+e_{r-4}
+\ldots)$}}
 \!\!\!\!\! \bra{e_{r-1}+e_{N}}{0}\nonumber \\
&+&\!\!\!\!\! \sum_{\foot{$Q+(e_N+e_{N-2}+\ldots)$}{$+(e_{r-3}+e_{r-5}
+\ldots)$}}
 \!\!\!\!\! q^{\frac{L}{2}}\bra{e_{r-2}+e_{N+1}}{0}\nonumber\\
&=&0
\end{eqnarray}
and is true due to lemma \ref{lemma2}b) with $a=N-r$. Therefore the
proof of recursion relation (\ref{recdownsmall})
is complete.

Finally the recursion relation for $r=n+1$ follows since due to lemma
\ref{lemma1} both sides of equation (\ref{recdowns}) are equal to
\begin{equation}
\sum_{\foot{$Q+(e_{N-1}+e_{N-3}+\ldots)$}{$+(e_{n-1}+e_{n-3}
+\ldots)$}}\!\!\!\!\bra{e_n+e_N}{0}.
\end{equation}
This concludes the proof of all recursion relations.

\section{Polynomial Expressions under the duality $q\rightarrow q^{-1}$}
\label{sec-duality}
\setcounter{equation}{0}

In this section we obtain the configuration sums of the RSOS model
in regime II by the duality transformation
$q\rightarrow q^{-1}$ from our results (\ref{result1})
and (\ref{result2}) in regime III. The RSOS model in regime II
corresponds to the coset conformal field theory
$\widehat{su}(2\nu-2)_1/\widehat{sp}(2\nu-2)_1$ \cite{Date}, \cite{Date1}.
We will also identify our
results with the branching functions of the $Z_{\nu-1}$-parafermion
model.

Under duality the Gaussian polynomials transform as follows
\begin{equation}
\label{Gaustrans}
\left[ \begin{array}{c} n \\ m \end{array} \right]_{q^{-1}}
=q^{m(m-n)}\left[ \begin{array}{c} n \\ m \end{array} \right]_{q}.
\end{equation}
Hence the object
\begin{equation}
X_L[A,u,Q,q]=\sum_Q q^{\frac{1}{4}mCm-\half Am}\prod_{i=1}^{\nu-2}
 \left[ \begin{array}{c} \half (Im+u+Le_N)_i \\ m_i \end{array} \right]_{q}
\end{equation}
transforms as
\begin{eqnarray}
\label{transformation}
X_L[A,u,Q,q^{-1}]=\sum_Q q^{\frac{1}{4}mCm
+\half (A-u)m-\half Lm_N}\prod_{i=1}^{\nu-2}
 \left[ \begin{array}{c} \half (Im+u+Le_N)_i \\ m_i \end{array} \right]_{q}.
\end{eqnarray}
To take the limit $L\rightarrow \infty$ we have to take out a
factor $q^{f(L)}$ where $f(L)$ is some
function of $L$. To determine the appropriate factor
we make the variable change to the dual variable $\n_i$
\begin{equation}
\label{def:n}
m_i+\n_i=\half (Im+u+Le_N)_i.
\end{equation}
This equation is equivalent to the following partition problem
\begin{equation}
\label{partition}
\sum_{i=1}^{\nu-1}i\n_i=\half (NL+\sum_{i=1}^{\nu-2}iu_i+(\nu-1)Q_{\nu-2})
\end{equation}
where $\n_i$ are non-negative integers ($i=1,2,\ldots,\nu-1$)
and $\n_{\nu-1}\equiv \half (m_{\nu-2}+Q_{\nu-2})$.

We can also solve equation (\ref{def:n}) for $m$ and obtain
\begin{equation}
\label{equ:m}
m=C^{-1}(u+Le_N-2\n)
\end{equation}
where $C^{-1}$ is the inverse of the Cartan matrix and is given by
\begin{equation}
\label{invCartan}
C^{-1}_{ij}=\left\{ \begin{array}{ll} \frac{(\nu-1-i)j}{\nu-1}
& {\rm for}\; j\leq i \\
 C^{-1}_{ji} & {\rm for}\; j>i \end{array} \right. .
\end{equation}
Hence (\ref{transformation}) reads in the dual variables $\n_i$
\begin{equation}
\label{trans}
q^{f(L)}X_L[A,u,Q,q^{-1}]
 =\sumt q^{\n C^{-1}\n-AC^{-1}\n-\frac{1}{4}uC^{-1}u
+\half AC^{-1}u}\prod_{i=1}^{\nu-2}
 \left[ \begin{array}{c} \n_i+m_i \\ \n_i \end{array} \right]
\end{equation}
where $m$ as in (\ref{equ:m}) and
$f(L)=\frac{1}{4}L^2N\frac{\nu-N-1}{\nu-1}+\half L(u-A)C^{-1}e_N$.
The sum $\sumt$ runs over all $\n_i\geq 0$ satisfying (\ref{partition}).
Notice that all the $L$ dependence in the exponent is independent of the
variable $\n$ and hence could be factored out.

Let us now calculate $\half L(u-A)C^{-1}e_N$ for the different cases.
Let us start with the ``down'' configuration
sums $X_L(s|r,r+N-2n)$. For the pair $(r,r+N-2n)$ to be admissable
it must fulfill $N+2\leq 2r+N-2n\leq 2\nu-N$ which
implies $N\leq \nu-r+n$. Hence we conclude that for the first term
in $X_L(s|r,r+N-2n)_{q^{-1}}$ (see (\ref{result1}) and (\ref{result2}))
\begin{eqnarray}
\half L(u-A)C^{-1}e_N&=&\half L( e_{\nu-r+n}+e_n)C^{-1}e_N\nonumber\\
&=&\frac{L}{2(\nu-1)}\left( (r-1-n)N+(\nu-1-N)n\right) .
\end{eqnarray}
Similarly for the second term in (\ref{result2})
\begin{eqnarray}
\half L(u-A)C^{-1}e_N+\frac{L}{2}&=&\half L(e_{\nu-r+n+1}
+e_{n-1})C^{-1}e_N+\frac{L}{2}\nonumber\\
&=&\frac{L}{2(\nu-1)}\left( (r-2-n)N+(\nu-1-N)(n-1)+\nu-1\right)\nonumber\\
&=&\frac{L}{2(\nu-1)}\left( (r-1-n)N+(\nu-1-N)n\right) .
\end{eqnarray}
Hence both terms have the same function $f$
\begin{equation}
f^{\rm down}(L)=\frac{1}{4}\frac{L^2}{\nu-1}N(\nu-1-N)
+\half \frac{L}{\nu-1}\left( (\nu-1-N)n+(r-n-1)N\right) .
\end{equation}
Similarly one can calculate the function $f$ for
the ``up'' character $X_L(s|r,r-N+2n)_{q^{-1}}$
\begin{equation}
f^{\rm up}(L)=\frac{1}{4}\frac{L^2}{\nu-1}N(\nu-1-N)
+\half \frac{L}{\nu-1}\left( (\nu-1-N)n+(\nu-r-n)N\right) .
\end{equation}

Hence if we define
\begin{equation}
\label{Xdual}
\tilde{X}_L[A,u,Q]=\sumt q^{\n C^{-1}\n-AC^{-1}\n
-\frac{1}{4}uC^{-1}u+\half AC^{-1}u}
 \prod_{i=1}^{\nu-2}\left[ \begin{array}{c} (\n+C^{-1}(u+Le_N-2\n))_i
\\ \n_i \end{array} \right]
\end{equation}
we have
\begin{eqnarray}
q^{f^{\rm down}(L)}X_L(s|r,r+N)_{q^{-1}}\!\!&=&\!\!
\tilde{X}_L[A^{\rm down}_{r,s},u^{\rm down}_{r,s},Q^{\rm down}_{r,s}]
\nonumber\\
q^{f^{\rm up}(L)}X_L(s|r,r-N)_{q^{-1}}\!\!&=&\!\!
\tilde{X}_L[A^{\rm up}_{r,s},u^{\rm up}_{r,s},Q^{\rm up}_{r,s}]
\end{eqnarray}
and
\begin{eqnarray}
q^{f^{\rm down}(L)}X_L(s|r,r+N-2n)_{q^{-1}}&\!\!
=&\!\!\tilde{X}_L[A^{\rm down}_{r-n,s},u^{\rm down}_{r-n,s}+e_n,
 Q^{\rm down}_{r-n,s}]\nonumber\\
 -\theta(r>n+1)&&\!\!\!\!\!\!\!\!\!\!
\tilde{X}_L[A^{\rm down}_{r-n-1,s},u^{\rm down}_{r-n-1,s}+e_{n-1},
 Q^{\rm down}_{r-n-1,s}]\nonumber\\
q^{f^{\rm up}(L)}X_L(s|r,r-N+2n)_{q^{-1}}&\!\!
=&\!\!\tilde{X}_L[A^{\rm up}_{r+n,s},u^{\rm up}_{r+n,s}+e_n,
 Q^{\rm up}_{r+n,s}]\nonumber\\
 -\theta(r<\nu-n)&&\!\!\!\!\!\!\!\!\!
\tilde{X}_L[A^{\rm up}_{r+n+1,s},u^{\rm up}_{r+n+1,s}
+e_{n-1},Q^{\rm up}_{r+n+1}]
\end{eqnarray}
for $n=1,2,\ldots,\left[ \frac{N}{2} \right]$.
$A^{\rm down}_{r,s}, A^{\rm up}_{r,s}, u^{\rm down}_{r,s}, u^{\rm up}_{r,s},
Q^{\rm down}_{r,s}$  and $Q^{\rm up}_{r,s}$ are
defined as in (\ref{down}) and (\ref{up}). The dependence
on $Q^{\rm down}_{r,s}$
and $Q^{\rm up}_{r,s}$ is only implicit through (\ref{partition}).
Since the dependence is only on the $(\nu-2)^{\rm th}$ component of
$Q$ we could drop the terms $(e_{n-1}+e_{n-3}+\ldots)$ and $(e_{n-2}
+e_{n-4}+\ldots)$.

Since $(C^{-1}e_N)_i\neq 0$ for
all $i\in \{1,2,\ldots,\nu-2\}$ all Gaussians in the product in (\ref{Xdual})
turn into $\frac{1}{(q)_{\n_i}}$ in the limit $L\rightarrow \infty$ and we get
\begin{eqnarray}
\label{Xinf}
\tilde{X}_{\infty}[A,u,Q]&\equiv&
\!\!\!\lim_{\begin{tabular}{c} {\scriptsize $L\rightarrow \infty$}
\\ {\scriptsize $ LN \equiv 0 \, {\rm mod}\,
 (2\nu-2)$}\end{tabular}}\!\! \tilde{X}_{L}[A,u,Q]\nonumber\\
 &=&\sumtt q^{\n C^{-1}\n-AC^{-1}\n-\frac{1}{4}uC^{-1}u
+\half AC^{-1}u} \prod_{i=1}^{\nu-2}\frac{1}{(q)_{\n_i}}.
\end{eqnarray}
where $\sumtt$ runs over $\n_i\geq 0$ such that
$\left( C^{-1}(\n-\half u)-\half Q\right)_{\nu-2}\in {\bf Z}$.
Notice that the $N$ dependence drops out.

According to \cite{Date} and \cite{Date1} the branching
functions $e_{jk}^{l}$ for the coset
model $\widehat{su}(2\nu-2)/\widehat{sp}(2\nu-2)$
satisfy
\begin{equation}
q^{\tilde{\eta}}e_{r-1,s-1}^{\nu-1}
=\lim_{\begin{tabular}{c} {\scriptsize $L\rightarrow \infty$}
\\ {\scriptsize $ LN \equiv 0 \, {\rm mod}\,
 (2\nu-2)$}\end{tabular}}q^{f^{\rm down}(L)}X_{L}(s|r,r+N)_{q^{-1}}
\end{equation}
where $\tilde{\eta}=\frac{1}{4(\nu-1)}
\left( \frac{\nu+1}{2}-r\right)^{2}-\frac{1}{4(\nu+1)}
\left( \frac{\nu+1}{2}-s\right)^2-\frac{1}{24}$.
Hence the fermionic representation for the branching functions is
\begin{equation}
\label{e_equ}
q^{\tilde{\eta}}e_{r-1,s-1}^{\nu-1}
=\tilde{X}_{\infty}[A_{r,s}^{\rm down},u_{r,s}^{\rm down},Q_{r,s}^{\rm down}]
\end{equation}
where $A^{\rm down}_{r,s}, u^{\rm down}_{r,s}$
and $Q^{\rm down}_{r,s}$ are defined as in (\ref{down}).

Finally, we would like to notice that (\ref{e_equ}) is also related to the
branching functions for the $Z_{\nu-1}$-parafermion model \cite{Lepowsky},
\cite{LepWil}, \cite{ZamFat}. The fermionic form of these branching
functions is given by \cite{Lepowsky}, \cite{KKMM}, \cite{KMM}, \cite{Foda}
\begin{equation}
\label{parafermion}
q^{\frac{c}{24}}q^{-\frac{(s-1)(\nu-s)}{2(\nu-1)(\nu+1)}}b_{2p-s+1}^{s-1}
 =\sum_{m_1,\ldots,m_{\nu-2}\geq 0, {\rm restr.}}
 q^{mC^{-1}m-e_{s-1}C^{-1}m}\prod_{i=1}^{\nu-2}\frac{1}{(q)_{m_i}}
\end{equation}
where the $m_i$'s are subject to the restriction
$(C^{-1}m)_{\nu-2}-\frac{1}{\nu-1}p\in {\bf Z}$ and $c=\frac{2(\nu-2)}{\nu+1}$
is the central charge. A bosonic form of the branching function
$b^l_m$ is given in terms of a double sum \cite{Kac}, \cite{JimboMiwa},
\cite{Distler}.
Identifying $p$ with
$\frac{(\nu-1)}{2}(C^{-1}u_{r,s}^{\rm down}+Q_{r,s}^{\rm down})_{\nu-2}$ and
taking out the factor
\begin{equation}
q^{-\frac{1}{4}u_{r,s}^{\rm down}C^{-1}u_{r,s}^{\rm down}
+\half A_{r,s}^{\rm down}C^{-1}u_{r,s}^{\rm down}}
=q^{\frac{1}{4(\nu-1)}((\nu-s)(s-1)-(r-1)(\nu-r))}
\end{equation}
establishes the relation between (\ref{e_equ}) and (\ref{parafermion}), namely
\begin{equation}
e^{\nu-1}_{r-1,s-1}=b^{s-1}_{r-1}.
\end{equation}

The branching functions of the $Z_{\nu-1}$-parafermion model correspond
to the coset models
$\widehat{su}(\nu-1)_1\times \widehat{su}(\nu-1)_1/\widehat{su}(\nu-1)_2$
\cite{Lepowsky}, \cite{LepWil}, \cite{ZamFat} \cite{Christe} and by level
rank duality \cite{Altschuler} to the ones of
$\widehat{su}(2)_{\nu}/U(1)$ \cite{Kac}, \cite{JimboMiwa}, \cite{Distler}.

\section*{Acknowledgements}

I would like to thank B. McCoy for suggesting
this problem to me and also for many helpful discussions and comments.
I would further like to thank A. Waldron and W. Orrick for discussions.
This work was partially supported by NSF grants PHY-9309888 and DMR-9404747.

\appendix
\section*{Appendix}
\section{Further Identities}
\label{ap-identities}
\setcounter{equation}{0}

In this appendix we give further useful identities in addition
to the ones already discussed in section \ref{sec-tele}.
Let us first give the telescopic expansion to the left of
length $M$ of the same type as (\ref{tele1})
\begin{eqnarray}
\sum_{Q} &&\!\! q^{-\cir{i+M}'}
\bra{{\cal A}_{< i}+{\cal A}_{\geq i+M}+2e_{i+M-1}}{{\cal B}_{<i}
+{\cal B}_{>i+M}+e_{i+M}}\nonumber \\
= \sum_{l=i}^{i+M} \, \sum_{Q}
&&\!\! q^{-\sum_{k=l}^{i+M}\cir{k}'} \bra{{\cal A}_{<i}
+{\cal A}_{\geq i+M}+2\sum_{k=l+1}^{i+M}e_k+2e_{i-1}\delta_{l,i}}
 {{\cal B}_{<i}+{\cal B}_{>i+M}+\sum_{k=l}^{i+M}e_k}
\end{eqnarray}
where we used a similar notation to (\ref{tele1}).
In the case that $i=1$ the term $2e_{i-1}\delta_{l,i}$ just drops out.

Next we give the telescopic expansion from left to right
of the same type as (\ref{tele2})
\begin{eqnarray}
\label{tele2r}
\sum_{Q} && q^{-\sum_{k=i}^{i+M}\cir{k}'}
\bra{{\cal A}_{<i}+{\cal A}_{\geq i+M}
+2\sum_{k=i}^{i+M}e_k}{{\cal B}_{<i}+{\cal B}_{>i+M}+\sum_{k=i}^{i+M}e_k}
\nonumber \\
=\sum_{l=i}^{i+M} \sum_{Q}
&&q^{-\sum_{k=l}^{i+M}\cir{k}'} \bra{{\cal A}_{<i}
+{\cal A}_{\geq i+M}+2\sum_{k=l+1}^{i+M}e_k}
 {{\cal B}_{<i}+{\cal B}_{>i+M}+\sum_{k=l}^{i+M}e_k}\nonumber \\
+\sum_{Q}&& q^{\half (\A{\geq i+M})_{i+M}}
 \bra{\A{<i} +\A{\geq i+M}}{\B{<i}+\B{>i+M}}.
\end{eqnarray}

The analogue of (\ref{teleend}) in opposite direction is given by
\begin{eqnarray}
\label{teleend1}
\sum_{Q}\,\sum_{l=2}^{i}q^{-\sum_{k=l}^{i}\cir{k}'}
\bra{\A{\geq i}+2\sum_{k=l+1}^{i}e_k+e_1}{\B{>i}+\sum_{k=l}^{i}e_k}\nonumber \\
=\sum_{Q}q^{-\cir{i}'-\sum_{k=1}^{i-2}\cir{k}'+\half}
\bra{\A{\geq i}+2\sum_{k=2}^{i-1}e_k+e_1}{\B{>i}+e_i+\sum_{k=1}^{i-2}}
\end{eqnarray}

Now we come to the telescopic expansions corresponding to
recursion relation (\ref{recursion'}). We first give the analogue
of (\ref{tele1}) (expansion to the right from $i$ to $i+M$)
\begin{eqnarray}
\label{tele1'}
\sum_{Q} &&\!\! q^{m_i-m_{i+1}}
\bra{{\cal A}+2e_{i+1}}{{\cal B}_{<i}+{\cal B}_{>i+M}+e_{i}}\nonumber \\
= \sum_{l=i}^{i+M} \,
\sum_{Q} &&\!\! q^{m_l-m_{l+1}\delta_{l,i+M}} \bra{{\cal A}
+2\sum_{k=i}^{l-1}e_k+2e_{i+M+1}\delta_{l,i+M}}
 {{\cal B}_{<i}+{\cal B}_{>i+M}+\sum_{k=i}^{l}e_k}.
\end{eqnarray}
Since the phase in (\ref{recursion'}) is independent of the
top entries ${\cal A}$ is just any vector with integer values.

The analogeous telescopic expansion to (\ref{tele2}) corresponding
to (\ref{recursion'}) is
\begin{eqnarray}
\label{tele2'}
\sum_{Q} &&\bra{{\cal A}+2\sum_{k=i}^{i+M}e_k}{{\cal B}_{<i}
+{\cal B}_{>i+M}+\sum_{k=i}^{i+M}e_k}\nonumber \\
=\sum_{l=i}^{i+M} \sum_{Q}
&&q^{m_{l}+1} \bra{{\cal A}+2\sum_{k=i}^{l-1}e_k}
 {{\cal B}_{<i}+{\cal B}_{>i+M}+\sum_{k=i}^{l}e_k}\nonumber \\
+\sum_{Q}&& \bra{\cal A}{\B{<i}+\B{>i+M}}.
\end{eqnarray}

The proof of (\ref{tele1'}) and (\ref{tele2'}) is analogeous
to the proofs of (\ref{tele1}) and (\ref{tele2})
respectively. Instead of using recursion relation (\ref{recursion})
one simply uses (\ref{recursion'}).
There are again analogeous formulas to (\ref{tele1'})
and (\ref{tele2'}) in the opposite direction the derivation
of which is left to the reader.

Next we have the mirror identity due to recursion relation (\ref{recursion'})
\begin{eqnarray}
\label{mirror'}
\sum_{l=i}^{M+i}\, \sum_{Q}&&\!\! q^{m_l}
\bra{{\cal A}+2\sum_{k=i}^{l-1}e_k}{{\cal B}_{<i}+{\cal B}_{>i+M}
+\sum_{k=i}^{l}e_{k}}\nonumber \\
= \sum_{l=i}^{i+M} \, \sum_{Q}
&&\!\! q^{m_l} \bra{{\cal A}+2\sum_{k=l+1}^{i+M}e_k}{{\cal B}_{<i}
+{\cal B}_{>i+M}+\sum_{k=l}^{i+M}e_k}.
\end{eqnarray}
The proof is again similar to the one for (\ref{mirror}).

Finally we have the extended mirror identity corresponding
to (\ref{recursion'})
\begin{eqnarray}
\label{exmirror'}
\sum_{l=i+2}^{M+i}\, \sum_{Q}&&\!\!\!\!\!\! q^{m_i-m_{i+1}+m_l}
\bra{{\cal A}+2e_{i+1}+2\sum_{k=l+1}^{i+M}e_k}{{\cal B}_{<i}
+{\cal B}_{>i+M}+e_i+\sum_{k=l}^{i+M}e_{k}}\nonumber \\
= \sum_{l=i}^{i+M-2} \, \sum_{Q}
&&\!\!\!\!\!\! q^{m_l+m_{i+M}-m_{i+M-1}} \bra{{\cal A}
+2e_{i+M-1}+2\sum_{k=i}^{l-1}e_k}{{\cal B}_{<i}
+{\cal B}_{>i+M}+e_{i+M}+\sum_{k=i}^{l}e_k}
\end{eqnarray}
the proof of which is similar to the one for (\ref{exmirror}).

For the proof of lemma \ref{lemma1} we need two further
identities. They are special telescopic expansions using recursion
relation (\ref{recursion'}).
\begin{eqnarray}
\label{special'}
\sum_{Q} q^{-\half m_1}\bra{{\cal A}+2e_1}{0}
=\sum_{Q} q^{\half m_1}
 \bra{{\cal A}+2\sum_{k=1}^{\nu-2}e_k}{\sum_{k=1}^{\nu-2}e_k}
\end{eqnarray}
where ${\cal A}$ is an arbitrary vector with integer
values and $Q\in \Ztwo$ arbitrary.
The derivation of this identity goes as follows.
Use recursion relation (\ref{recursion'}) in the first
slot of the lhs of (\ref{special'})
\begin{eqnarray}
&&\sum_{Q} q^{\half m_1} \bra{\cal A}{0}+\sum_{Q}
q^{-\half m_1} \bra{\cal A}{-e_1}\nonumber \\
&=&\sum_{Q} q^{\half m_1} \bra{\cal A}{0}+\sum_{Q}
q^{\frac{3}{2} m_1-m_2+1} \bra{{\cal A}+2e_2}{e_1}
\end{eqnarray}
where we made the variable change $m_1\rightarrow m_1+2$ in the
second term to get the second line. We can now use (\ref{tele1'})
with $i=1$ and
$i+M=\nu-2$ on the second term (the extra exponent $\half m_1$
does not matter since one uses only variable changes for $m_2$ and higher
in the derivation of (\ref{tele1'})) and obtain
\begin{equation}
q^{\half m_1}\sum_{Q}\left( \bra{\cal A}{0}
+\sum_{l=1}^{\nu-2}q^{m_l+1}\bra{{\cal A}
+2\sum_{k=1}^{l-1}e_k}{\sum_{k=1}^{k=l}e_k}\right)
\end{equation}
which is the rhs of (\ref{special'}) via (\ref{tele2'}).

And last but not least we have the following identity
\begin{eqnarray}
\label{special1'}
\sum_{Q}q^{\half m_1+m_i-m_{i-1}}\bra{{\cal A}
+2\sum_{k=1}^{i-1}e_k}{\B{>i}+\sum_{k=1}^{i-2}e_k+e_{i}}\nonumber\\
=\sum_{l=2}^{i}\sum_{Q}q^{-\half m_1+m_l}\bra{{\cal A}
+2e_1+2\sum_{k=l+1}^{i}e_k}{\B{>i}+\sum_{k=l}^{i}e_k}.
\end{eqnarray}
For the derivation of this identity we use recursion
relation (\ref{recursion'}) in the first slot on the rhs
of (\ref{special1'}) for
$l>2$.
\begin{eqnarray}
\label{pr:special1'}
q^{\half m_1}\sum_{Q}&&\left( q^{-m_1+m_2}\bra{{\cal A}
+2e_1+2\sum_{k=3}^{i}e_k}{\B{>i}+\sum_{k=2}^{i}e_k}\right. \nonumber\\
&&+\sum_{l=3}^{i}q^{m_l}\bra{{\cal A}
+2\sum_{k=l+1}^{i}e_k}{\B{>i}+\sum_{k=l}^{i}e_k}\nonumber\\
&&+\left. \sum_{l=3}^{i}q^{m_1-m_2+1+m_l}\bra{{\cal A}
+2e_2+2\sum_{k=l+1}^{i}e_k}{\B{<i}+e_1+\sum_{k=l}^{i}e_k}\right)
\end{eqnarray}
We can combine the first two terms via the analogue
of (\ref{tele1'}) in the opposite direction. In the last sum
we can proceed analogeously to the
derivation of the extended mirror identity corresponding to
recursion relation (\ref{recursion'}): expand the $l^{\rm th}$
term from $1$ to $l-2$ which leads to another summation over $l'$.
Swapping the two sums and recombining the $l$ sum yields for
(\ref{pr:special1'})
\begin{eqnarray}
q^{\half m_1}\sum_{Q}&&\left( q^{-m_{i-1}+m_i}\bra{{\cal A}
+2e_{i-1}}{\B{>i}+e_i}\right. \nonumber\\
&&+\left. \sum_{l'=1}^{i-2}q^{m_i-m_{i-1}+1+m_{l'}}\bra{{\cal A}
+2\sum_{k=1}^{l'-1}e_k+2e_{i-1}}{\B{>i}+\sum_{k=1}^{l'}e_k+e_i}\right)
\end{eqnarray}
which is the lhs of (\ref{special1'}) due to (\ref{tele2'}).

Again analogous formulas to (\ref{special'}) and (\ref{special1'})
hold in opposite direction. The derivation of these is
left to the reader.

\section{Proof of the Lemmas}
\label{ap-lemmas}
\setcounter{equation}{0}

In this appendix we give the proofs of the three lemmas stated
in section \ref{sec-tele}. We begin with the
proof of lemma \ref{lemma1}.\newline
{\bf Proof of lemma \ref{lemma1}}

We first consider the case that $s$ is odd. The lemma is trivially
true for $s=1$. For $s\geq 3$ we
make the variable change $m_i\rightarrow m_i+1$ for
$i=1,3,\ldots,s-2$ and get for the lhs of (\ref{equ:lemma1})
\begin{eqnarray}
\label{pr:lemma1}
\lefteqn{\sum_{Q+(e_{s-2}+e_{s-4}+\ldots)} q^{-\half m_{s-1}}
 \bra{{\cal A} +e_{s-1}}{0}}\nonumber \\
&&=\sum_{Q} q^{-\cir{1}-\cir{3}-\ldots-\cir{s-2}-\half m_{s-1}+\frac{s-1}{4}}
 \bra{{\cal A}+2e_2+2e_4+\ldots+2e_{s-1}}{e_1+e_3+\ldots+e_{s-2}}.
\end{eqnarray}
Next notice that we can rewrite the exponent as
\begin{equation}
-\cir{1}-\cir{3}-\ldots-\cir{s-2}-\half m_{s-1}=m_1-m_2
+m_3-m_4\pm\ldots+m_{s-2}-m_{s-1}.
\end{equation}
Now using (\ref{tele1'}) with $i=s-2$ and $i+M=\nu-2$, (\ref{mirror'}),
$\frac{s-3}{2}$ times (\ref{exmirror'})
and the analogue of (\ref{tele1'}) in the opposite direction
we obtain for (\ref{pr:lemma1})
\begin{eqnarray}
\sum_{Q}q^{-\cir{\nu-2}-\cir{\nu-4}-\ldots-\cir{\nu-s+1}
-\half m_{\nu-s}+\frac{s-1}{4}}
 \bra{{\cal A}+2e_{\nu-s}+2e_{\nu-s+2}+\ldots+2e_{\nu-3}}{e_{\nu-s+1}
+e_{\nu-s+3}+\ldots+e_{\nu-2}}
\end{eqnarray}
which in turn is the rhs of (\ref{equ:lemma1}) after making
the variable change $m_i\rightarrow m_i-1$ for
$i=\nu-s+1,\nu-s+3,\ldots,\nu-2$.

We proceed similarly for even $s$.
For $s=2$ use (\ref{special'}) with $\tilde{\cal A}={\cal A}-e_1$
and make the variable change $m_i\rightarrow m_i-1$ for
$i=1,2,\ldots,\nu-2$.
For $s>2$ we make the variable change $m_i\rightarrow m_i+1$ for
$i=2,4,\ldots,s-2$ and obtain for the lhs of (\ref{equ:lemma1})
\begin{equation}
\label{pr:lemma1'}
\sum_{Q} q^{-\cir{2}-\cir{4}-\ldots-\cir{s-2}-\half m_{s-1}+\frac{s-2}{4}}
 \bra{{\cal A}+e_1+2e_3+2e_5+\ldots+2e_{s-1}}{e_2+e_4+\ldots+e_{s-2}}.
\end{equation}
The exponent can again be rewritten as
\begin{equation}
-\cir{2}-\cir{4}-\ldots-\cir{s-2}-\half m_{s-1}=-\frac{m_1}{2}+m_2-m_3
+m_4-m_5\pm\ldots+m_{s-2}-m_{s-1}.
\end{equation}
Using (\ref{tele1'}) with $i=s-2$ and $i+M=\nu-2$, (\ref{mirror'}),
$\frac{s-4}{2}$ times (\ref{exmirror'})
and (\ref{special1'}) we obtain for (\ref{pr:lemma1'})
\begin{eqnarray}
\!\!\!\!\!\!\lefteqn{\sum_{Q} q^{\half m_1+m_{\nu-2}-m_{\nu-3}\pm\ldots
+m_{\nu-s+2}-m_{\nu-s+1}+\frac{s-2}{4}}}\nonumber\\
 &&\bra{{\cal A}+e_1+2e_2+2e_3+\ldots+2e_{\nu-s+1}+2e_{\nu-s+3}
+\ldots+2e_{\nu-3}}
 {e_1+e_2+e_3+\ldots+e_{\nu-s}+e_{\nu-s+2}+\ldots+e_{\nu-2}}\nonumber \\
\!\!\!\!\!\!\!\lefteqn{=\sum_{Q} q^{-\cir{\nu-2}-\cir{\nu-4}-\ldots
-\cir{\nu-s}-\cir{\nu-s-1}-\ldots-\cir{1}-\half m_{\nu-s}
+\frac{s-2}{4}}}\nonumber \\
 &&\bra{{\cal A}+e_1+2e_2+2e_3+\ldots+2e_{\nu-s+1}+2e_{\nu-s+3}
+\ldots+2e_{\nu-3}}
 {e_1+e_2+e_3+\ldots+e_{\nu-s}+e_{\nu-s+2}+\ldots+e_{\nu-2}}
\end{eqnarray}
which is the rhs of (\ref{equ:lemma1}) after making the variable
change $m_i\rightarrow m_i-1$ for
$i=1,2,3,\ldots,\nu-s,\nu-s+2,\ldots,\nu-2$.
\newline
{\bf Proof of lemma \ref{lemma2}a)}

For the proof of this lemma we need to distinguish between
$\bar{a}=N-a$ even and odd. Let us start with the case
that $\bar{a}$ even. We make the variable change
$m_i\rightarrow m_i+1$ for $i=a+1,a+3,\ldots,N-1$ on the lhs of
(\ref{equ:lemma2a}) and get
\begin{eqnarray}
\!\!\!\!\!\lefteqn{\sum_{Q+(e_{N-1}+e_{N-3}+\ldots)}
\!\!\! q^{-\cir{a+1}'-\cir{a+3}'-\ldots-\cir{N-1}'
+\frac{\bar{a}}{4}}}\nonumber\\
&&\:\:\:\bra{\B{\geq N}+2e_a+2e_{a+2}+\ldots+2e_{N-2}+e_N}{e_{a+1}
+e_{a+3}+\ldots+e_{N-1}}
\end{eqnarray}
We can now use the telescopic expansion (\ref{tele1}) to
the left from $a+1$ to 1, flip the expansion via the mirror
identity (\ref{mirror})
and apply the extended mirror identity (\ref{exmirror}). Repeat
this $\frac{\bar{a}-2}{2}$ times and recombine via the telescopic
expansion (\ref{tele2r})
which yields
\begin{eqnarray}
\!\!\!\!\!\!\!\!\lefteqn{\sum_{Q+(e_{N-1}+e_{N-3}+\ldots)}\!\!\!
q^{-\cir{1}'-\cir{3}'-\ldots-\cir{\bar{a}-1}'-\cir{\bar{a}}'
 -\ldots-\cir{N-1}'+\frac{\bar{a}}{4}}}\nonumber\\
&&\:\:\bra{\B{\geq N}+2e_2+2e_{4}+\ldots+2e_{\bar{a}-2}+2e_{\bar{a}-1}
+\ldots+2e_{N-1}+e_N}{e_{1}+e_{3}+\ldots+e_{\bar{a}-1}
 +e_{\bar{a}}+\ldots+e_{N-1}}\nonumber\\
\!\!\!\!\!\!\lefteqn{-\sum_{Q+(e_{N-1}+e_{N-3}+\ldots)}\!\!\!
q^{-\cir{1}'-\cir{3}'-\ldots-\cir{\bar{a}-3}'+\frac{\bar{a}}{4}}}\nonumber\\
&&\:\:\bra{\B{\geq N}+2e_2+2e_4+\ldots+2e_{\bar{a}-2}+e_N}{e_1
+e_3+\ldots+e_{\bar{a}-3}}\nonumber\\
\!\!\!\!\!\!\lefteqn{=\sum_{Q+a\rho_{<N}+(e_{\bar{a}}+e_{\bar{a}+2}
+\ldots)_{<N}}\bra{\B{\geq N}+e_{\bar{a}-1}+e_{N-1}}{0}}\nonumber\\
\!\!\!\!\!\!\lefteqn{-\sum_{Q+a\rho_{<N}+(e_{\bar{a}-1}+e_{\bar{a}+1}
+\ldots)_{<N}}q^{\half}\bra{\B{\geq N}+e_{\bar{a}-2}+e_N}{0}}
\end{eqnarray}
where we got the equal sign by making the variable changes
$m_i\rightarrow m_i-1$ for $i=1,3,\ldots,\bar{a}-1,\bar{a},\ldots,N-1$
for the first term and $i=1,3,\ldots,\bar{a}-3$ for the second term.

The proof for $\bar{a}$ odd is very similar. This time we make the
variable change $m_i\rightarrow m_i+1$ for $i=1,2,\ldots,a,a+2,\ldots,
N-1$ on the lhs of (\ref{equ:lemma2a}) and get
\begin{eqnarray}
\!\!\!\!\!\!\!\!\!\!\!\!\lefteqn{\sum_{Q+(e_{N-1}+e_{N-3}+\ldots)}\!\!\!
 q^{-\cir{1}'-\cir{2}'-\ldots-\cir{a}'-\cir{a+2}'-\ldots-\cir{N-1}'
+\frac{\bar{a}+1}{4}}}\nonumber\\
&&\bra{\B{\geq N}+e_1+2e_2+2e_3+\ldots+2e_{a+1}+2e_{a+3}+\ldots
+2e_{N-3}+e_N}{e_1+e_2+\ldots+e_a+e_{a+2}+\ldots+e_{N-1}}
\end{eqnarray}
Instead of using (\ref{tele1}) one can now apply the special
telescopic expansion (\ref{teleend1}). After that one proceeds exactly
as before:
flip the expansion via the mirror identity (\ref{mirror})
followed by the extended mirror identity (\ref{exmirror})
$\frac{\bar{a}-3}{2}$ times
and recombine via (\ref{tele2})
\begin{eqnarray}
\!\!\!\!\!\!\!\!\lefteqn{\sum_{Q+(e_{N-1}+e_{N-3}+\ldots)}\!\!\!
q^{-\cir{2}'-\cir{4}'-\ldots-\cir{\bar{a}-1}'-\cir{\bar{a}}'
 -\ldots-\cir{N-1}'+\frac{\bar{a}-1}{4}}}\nonumber\\
&&\:\bra{\B{\geq N}+e_1+2e_3+2e_{5}+\ldots+2e_{\bar{a}-2}+2e_{\bar{a}-1}
+\ldots+2e_{N-1}+e_N}{e_{2}+e_{4}+\ldots+e_{\bar{a}-1}
 +e_{\bar{a}}+\ldots+e_{N-1}}\nonumber\\
\!\!\!\!\!\!\!\lefteqn{-\sum_{Q+(e_{N-1}+e_{N-3}+\ldots)}\!\!\!
q^{-\cir{2}'-\cir{4}'-\ldots-\cir{\bar{a}-3}'+\frac{\bar{a}-1}{4}}}\nonumber\\
&&\:\:\bra{\B{\geq N}+e_1+2e_3+2e_5+\ldots+2e_{\bar{a}-2}+e_N}{e_2+e_4
+\ldots+e_{\bar{a}-3}}
\end{eqnarray}
Making the variable change $m_i\rightarrow m_i-1$ for
$i=2,4,\ldots,\bar{a}-1,\bar{a},\ldots,N-1$ for
the first term and $i=2,4,\ldots,\bar{a}-3$ for the second
term gives exactly the rhs of (\ref{equ:lemma2a}).
\newline
{\bf Proof of lemma \ref{lemma2}b}

The proof is similar to the proof of part a). We start this time with
the case $\bar{a}$ odd. Making the variable change
$m_i\rightarrow m_i+1$ for $i=a+2,a+4,\ldots,N-1$ we get for the lhs
of (\ref{equ:lemma2b})
\begin{eqnarray}
\!\!\!\!\!\!\lefteqn{\sum_{Q+(e_{N-1}+e_{N-3}+\ldots)}\!\!\! q^{-\cir{a+2}'
-\cir{a+4}'-\ldots-\cir{N-1}'+\frac{\bar{a}-1}{4}}}\nonumber\\
&&\:\:\bra{\B{>N}+2e_{a+1}+2e_{a+3}+\ldots+2e_{N-2}+e_N}{e_{a+2}+e_{a+4}
+\ldots+e_{N-1}}
\end{eqnarray}
Again we telescopically expand to the left from $a+2$ to 1, use the mirror
identity (\ref{mirror}) followed by
the extended mirror identity (\ref{exmirror}) $\frac{\bar{a}-3}{2}$ times.
The only difference to part a) is that we now
recombine the expansion via (\ref{tele1}) instead of (\ref{tele2}) with
$i=\bar{a}-2$ and $i+M=N$
\begin{eqnarray}
\!\!\!\!\lefteqn{\sum_{Q+(e_{N-1}+e_{N-3}+\ldots)}\!\!\! q^{-\cir{1}'-\cir{3}'
-\ldots-\cir{\bar{a}-2}'+\frac{\bar{a}-1}{4}}}\nonumber\\
&&\bra{\B{>N}+2e_2+2e_{4}+\ldots+2e_{\bar{a}-1}+e_N}{e_{1}+e_{3}+\ldots
+e_{\bar{a}-2}}\nonumber\\
\!\!\!\!\lefteqn{-\sum_{Q+(e_{N-1}+e_{N-3}+\ldots)}\!\!\! q^{-\cir{1}'-\cir{3}'
-\ldots-\cir{\bar{a}-2}'-\cir{\bar{a}-1}'-
 \ldots-\cir{N}'+\frac{L+1}{2}+\frac{\bar{a}-1}{4}}}\nonumber\\
&&\bra{\B{>N}+2e_2+2e_4+\ldots+2e_{\bar{a}-3}+2e_{\bar{a}-2}+\ldots+2e_{N-1}
+e_N+2e_{N+1}}{e_1+e_3+\ldots+e_{\bar{a}-2}+e_{\bar{a}-1}+
 \ldots+e_N}
\end{eqnarray}
where the second term is the term $l=i+M=N$ in (\ref{tele1}) and the
extra phase $\frac{L+1}{2}$ comes from the fact that there is the entry
$\half(L+1)$ in the $N^{\rm th}$ slot.
Changing variables $m_i\rightarrow m_i-1$ for $i=1,3,\ldots,\bar{a}-2$
for the first term and $i=1,3,\ldots,\bar{a}-2,\bar{a}-1,\ldots,N$
for the second term yields the rhs of (\ref{equ:lemma2b}).

The case $\bar{a}$ even goes the same with the only difference
that the first expansion is via the special telescopic
expansion (\ref{teleend1})
instead of (\ref{tele1}).
\newline
{\bf Proof of lemma \ref{lemma3}a}

The proof of this lemma is analogous to the proof of
lemma \ref{lemma2}a). We distinguish between $r-\bar{a}-1$ even and odd.
For $r-\bar{a}-1$ even one changes variables $m_i\rightarrow m_i+1$
for $i=N+1,N+3,\ldots,r+a$ for the lhs of (\ref{equ:lemma3a}). Then
once again one
telescopically expands this time to the right from $r+a$ to $\nu-2$,
uses the mirror identity followed by the extended mirror identity
$\frac{r-\bar{a}-1}{2}$ times and recombines via (\ref{tele2}).
The appropriate variable changes then yield the rhs of (\ref{equ:lemma3a}).

For $r-\bar{a}-1$ odd again telescopically expand to the right
from $r+a$ to $\nu-2$, use the mirror identity followed by the extended mirror
identity $\frac{r-\bar{a}-2}{2}$ times and recombine via (\ref{tele1}).
The appropriate variable changes then yield the rhs of (\ref{equ:lemma3a}).
\newline
{\bf Proof of lemma \ref{lemma3}b}

Again we first consider $r-\bar{a}-1$ even. Change variables
$m_i\rightarrow m_i+1$ for $i=N,N+2,\ldots,r+a-1$ on the lhs of
(\ref{equ:lemma3b}), telescopically expand to the left from $r+a-1$
to $\nu-2$, use the mirror identity and the extended mirror
identity $\frac{r-\bar{a}-1}{2}$ times and combine via (\ref{tele2}).

For $r-\bar{a}-1$ odd change variables $m_i\rightarrow m_i+1$ for
$i=N+1,N+3,\ldots,r+a-1$ on the lhs of (\ref{equ:lemma3b}), telescopically
expand to the left from $r+a-1$ to $\nu-2$, use the mirror identity and
the extended mirror identity $\frac{r-\bar{a}-2}{2}$ times and combine
via (\ref{tele1}).

\end{document}